\documentclass[preprint,superscriptaddress,double-spaced,floatfix,showkeys,showpacs,longbibliography]{revtex4}
\usepackage{graphicx,amsmath,bbm,bm,array,subfigure}
\usepackage{appendix}
\usepackage{lipsum}
\usepackage{dsfont}
\usepackage{calrsfs}
\usepackage{lineno}
\usepackage{bbold}
\usepackage[linktocpage=true,colorlinks=true,linkcolor=blue,citecolor=blue,urlcolor=blue]{hyperref}
\usepackage{MnSymbol}
\def\nn{\nonumber\\}
\newcommand\Snn{\mathcal{S}_{nn}(\vec{k},\omega)}

\begin{document}
\title {Role of slow, out-of-equilibrium modes on the dynamic 
structure factor near the QCD critical point}
\author{Golam Sarwar}
\email{golamsarwar1990@gmail.com}
\affiliation{Theory Division, Physical Research Laboratory, Navrangpura, Ahmedabad 380 009, India}
\author{Md Hasanujjaman}
\email{jaman.mdh@gmail.com}
\affiliation{Department of Physics, Darjeeling Government College, Darjeeling- 734101, India}
\author{Jan-e Alam}
\email{jane@vecc.gov.in}
\affiliation{Variable Energy Cyclotron Centre, 1/AF Bidhan Nagar, Kolkata- 700064, India}
\affiliation{Homi Bhabha National Institute, Training School Complex, Mumbai - 400085, India}
\def\zbf#1{{\bf {#1}}}
\def\bfm#1{\mbox{\boldmath $#1$}}
\def\hf{\frac{1}{2}}
\def\sl{\hspace{-0.15cm}/}
\def\omit#1{_{\!\rlap{$\scriptscriptstyle \backslash$}
{\scriptscriptstyle #1}}}
\def\vec#1{\mathchoice
{\mbox{\boldmath $#1$}}
{\mbox{\boldmath $#1$}}
{\mbox{\boldmath $\scriptstyle #1$}}
{\mbox{\boldmath $\scriptscriptstyle #1$}}
}
\def \beq{\begin{equation}}
\def \eeq{\end{equation}}
\def \beqa{\begin{eqnarray}}
\def \eeqa{\end{eqnarray}}
\def \pd{\partial}
\def \nn{\nonumber}
\begin{abstract}
The role of slow out of equilibrium modes (OEM), introduced to extend the validity of 
hydrodynamics near the QCD critical point on the power spectrum of  dynamical density 
fluctuations has been studied. We have used the equation of motion of slow modes for 
the situation when the extensive nature of thermodynamics is not altered due to the 
introduction of OEM. We find that the extensivity condition puts an extra constraint 
on the coupling of OEM  with the four divergence of velocity. The dynamic structure factor 
($\Snn$) in presence of the OEM  shows four Lorentzian peaks asymmetrically positioned 
about $\omega (\text{frequency})=0$, whereas the dynamic structure factor without the 
presence of any OEM shows three well-known Lorentzian peaks. The width of the peaks
are  reduced in presence of OEM. We find that the asymmetric peaks originate due to 
the coupling of the out-of-equilibrium modes with the hydrodynamic modes. It is
also shown that the OEM has negligible effects on $\Snn$ if first order hydrodynamics 
(relativistic Navier-Stokes) is used.  The introduction of OEM reduces the width of the 
Rayleigh peak indicating the reduction in the decay rate of the fluctuation which leads 
to slowing down, a well-known characteristics of the critical end point (CEP). 
\end{abstract}

\maketitle
\section{Introduction}
One of the outstanding issue in Relativistic High Energy Collision 
Experiments (RHIC-E) is to detect the Critical End Point (CEP) 
located at some critical baryonic chemical potential
($\mu_c$) and critical temperature ($T_c$) 
in the QCD (Quantum Chromodynamics) phase diagram. 
The existence of the CEP in QCD was suggested in 
Refs.~\cite{Halasz:1998qr,PhysRevD.49.426,PhysRevD.42.1757,
BERGES1999215,PhysRevD.62.105008} 
based on the effective field theoretic models
and by lattice QCD simulations  ~\cite{FODOR200287,Fodor:2004nz}. 
The experimental search for the CEP has been taken up through the beam energy 
scan (BES) program at 
the Relativistic Heavy Ion Collider (RHIC)~\cite{STAR:2010vob}. 
The search will continue in future experiments 
at the Facility for Anti-proton and Ion Research (FAIR) and Nuclotron- based Ion Collider fAcility (NICA). 
The exact location of the CEP is not known from the first 
principle because of the difficulties associated 
with the sign problem of the spin 1/2 Dirac particle (quark) in lattice 
QCD calculations~\cite{Gavai:2014ela}. Some of the QCD based 
effective models such as NJL and PNJL predict the location of the CEP
~\cite{MASAYUKI1989668,PhysRevD.41.1610,PhysRevD.49.426,PhysRevD.62.105008} 
but the results are dependent on the parameters of the models.
However, the prediction of the location of the CEP is not the only problem.
Even if the location of the CEP is predicted accurately
its experimental detection is extremely challenging
because the measured quantities are obtained by integration over
the space-time evolution history of the fireball produced
in relativistic heavy ion collisions, that is, the  experimental
results are superposition of all temperatures and 
densities through which the system  passes, not only from the 
single point ($\mu_c$,\,$T_c$) of the phase 
diagram.  The space time evolution of the fireball 
are studied by solving relativistic hydrodynamic equations
controlling the conservation of the density of conserved 
quantities. The hydrodynamic modeling has been used to study the evolution 
of the fireball created in relativistic nuclear collisions with the 
inclusion of the CEP~\cite{Lupeidu2021,Bleicher:2012qve,Aguiar2007,Nonaka}. 
However, the hydrodynamics breaks down near the  CEP
~\cite{Stephanov:2017ghc} due to long range correlations and 
enhanced fluctuations~\cite{AsakawaPhysRevLett.85.2072,Jeon2003eventbyevent,
Koch2008hadronic}. The CEP causes the system to linger its relaxation to 
equilibrium, termed as the `critical slowing 
down'~\cite{Berdnikov:1999ph,Stanley,SKMa}. It has been shown in
Ref.~\cite{Stephanov:2017ghc} that the validity of the hydrodynamics can be extended 
near CEP by introducing a scalar non-hydrodynamic field. 


The hydrodynamics is used to describe the slowly evolving
modes of the macroscopic system while the faster     
non-hydrodynamic modes are set by the collision dynamics at the 
microscopic level. The hydrodynamics can be applied to systems in local 
equilibrium. The time required to achieve local equilibrium
is much shorter than the time required to attain global 
equilibrium and this separation of time scale permits the application of hydrodynamics. 
At the CEP the correlation length diverges leading to the
divergence of the time scale for local equilibration which leads to
the break down of the hydrodynamics.
In other words, if the system encounters CEP then the correlation 
length ($\xi$) diverges and the relaxation time
which evolves as $\sim \xi^3$, also diverges leading to critical slowing down. 
Consequently, the system stay away from local thermal equilibrium 
and the hydrodynamics becomes inapplicable.  
In such situation, a new variable, $\phi$
representing non-hydrodynamic mode is included in the definition
of entropy along with other hydrodynamical variables
to extend the validity of the hydrodynamics in the vicinity of the CEP
~\cite{Stephanov:2017ghc}. The slow out-of-equilibrium modes (OEM)
are not treated 
separately~\cite{Stephanov:2017ghc,Rajagopal2020}
as they are coupled with hydrodynamic modes.  
In this regards, it is very important to understand the role of $\phi$ on the critical point dynamics. 
In the present work the effect of $\phi$ on the spectral structure will be studied 
using relativistic causal dissipative hydrodynamics proposed by Israel and Stewart 
(IS)~\cite{Israel:1976tn}. 
The dispersion relations for the hydrodynamic
modes are governed by the set of hydrodynamic equations. 
The extensivity of thermodynamics restricts the coupling of 
slow modes with gradients of other hydrodynamic fields.

The fluctuations in condensed matter system has been extensively studied. 
Specifically, the  power spectrum of the correlation of density fluctuations 
or the spectral structure ($\Snn$) has
been studied both experimentally and theoretically  and 
shown to be crucially dependent on transport coefficients  
~\cite{Stanley, SKMA2018,Linda}. 
The thermally excited fluctuations are governed by the 
the transport coefficients of the medium according to
the Onsager's hypothesis~\cite{OnsagarPhysRev.37.405}. 
The light and neutron scatterings experiments have been performed to investigate the 
properties of fluctuations experimentally in condensed matter physics. 
The spectrum of scattered light contains separately identifiable peaks of Lorentzian distribution, 
called Rayleigh (R) peak~\cite{Rayleigh1881} at angular frequency, $\omega=0$, and Brillouin (B) peaks located symmetrically
in the opposite side of R-peak, experimentally detected by Fleury and Boon~\cite{FlerryandBoon1969}. 
The R-line arises from the entropy or temperature fluctuations 
at constant pressure, whereas B-lines arise from the 
pressure fluctuations at constant entropy. 
The width of the R-line is connected to $\kappa/(\rho C_{P})$~\cite{Stanley}, 
where $\kappa$ is the thermal conductivity, $C_{P}$ is the isobaric specific heat
of the fluid with mass density, $\rho$. 
Therefore, if the order of divergence for $\kappa$ is weaker than $C_{P}$, 
then a narrow R-line appears. Since the width represents the decay rate of the
fluctuation a narrow width will indicate the slower decay of fluctuation.
Two Brillouin-peaks are positioned at  $\pm c_{s}k$ 
with respect to the frequency of the incident light, where $c_s$
is the speed of sound and $k$ is the wave vector. The finite width of the 
B-line provides information about the 
values of transport coefficients such as shear, and bulk viscosities. 
The ratio of intensities of R-line to B-line is: 
$I_{R}/2I_{B}=C_{P}/C_{V}-1=K_{T}/K_{S}-1$, called the Landau-Placzek ratio
where $K_T$ and $K_S$ are isothermal and isothermal and adiabatic 
compressibilities respectively.
Therefore, the study of the structure factor is very useful
to understand the behaviour of the fluid near the CEP
as some of the transport coefficients and response functions
change drastically. Near the CEP the B-peaks tend to vanish
giving rise to severe modification in $\Snn$.  
In QCD no external probe exists to measure such
drastic change in $\Snn$.
However, it is a well-known feature of the critical phenomena that 
the behaviour of a large class of systems near CEP are independent 
of the dynamical details.
Since, the CEP in QCD belongs to same universality class, 
$\mathcal{O}(4)$~\cite{Halasz:1998qr,BERGES1999215}, as that of liquid-gas critical point, 
the role of such slow modes can be tested in liquid-gas system, 
and the knowledge can be very helpful to guide the theoretical modeling relevant to the CEP 
of QCD. In  this context the qualitative and 
quantitative effects of the slow modes on various physical quantities 
should be studied.  In the present work the effects of these modes on
the spectral structure of the fluctuations near the CEP 
~\cite{10.1143/PTP.122.881,Minami,Hasan2} has been investigated. 
More specifically, we will study role of  the slow modes 
on the Rayleigh and Brillouin peaks of $\Snn$.   
Thermal fluctuations can cause entropy production  leading  to 
fluctuation in multiplicity.  However, their ensemble average is 
unaffected~\cite{Nagai2016}. Thermal fluctuations may not affect lower flow harmonics 
but affect their correlation. Recently, CMS collaborations have measured higher 
flow harmonics which again carry a strong signature of thermal 
fluctuations~\cite{CMS:2013jlh} and hence that of CEP.

The paper is organized as follows. In the next section the expression for $\Snn$ will be derived
by including the variable, $\phi$ representing the non-hydrodynamic mode.
In Sec.\ref{sec3} the results are presented and Sec.\ref{sec4} is devoted 
to summary and discussion.

\section{Hydrodynamics with out-of-equilibrium modes}
\label{sec2}
The evolution of fluid near the CEP within the scope of  fluid dynamics 
can be extended by introducing an extra slow degree of freedom
representing soft mode~\cite{Stephanov:2017ghc}. 
Here we use the IS hydrodynamical model~\cite{Israel:1976tn} in 
Eckart frame~\cite{Eckart:1940te} to study the effects 
of this slow degrees of freedom. The signature metric for the Minkowski 
space time is taken as, $g^{\mu\nu}=(-1, \,1,\, 1\,, 1)$. The fluid velocity field, 
$u^{\mu}$ is normalized as $u^{\mu}u_{\mu}=-1$. 
The energy-momentum and charge conservation equations are:
\begin{equation}
\partial_{\mu} T^{\mu \nu}=0
\end{equation}
 and
 \begin{equation}
     \partial_{\mu} J^{\mu}=0.
 \end{equation}
 where,
 \begin{equation}
T^{\mu\nu}=\epsilon u^{\mu}u^{\nu}+(P+\Pi)\Delta^{\mu\nu}+h^{\mu}u^{\nu}+u^{\mu}h^{\nu}+\pi^{\mu\nu},
 \end{equation}
 
 \begin{equation}
     J^{\mu}=nu^{\mu}+n^{\mu}\,,
 \end{equation}
where  $q^{\mu}=h^{\mu}-n^{\mu}(\epsilon+P)/n$ is the heat flux, 
$h^{\mu}$ is the vector dissipation or dissipative energy flow, 
$u^{\mu}$ is the flow velocity, 
with $u_{\mu}h^{\mu}=u_{\mu}\pi^{\mu\nu}=0$,
$n$ is conserved charge number density (net baryon density for RHIC-E), 
$n^{\mu}$ is dissipative current density, $\epsilon$ is energy density, 
$P$ is the pressure, $\Pi$ is the scalar dissipation or the bulk-stress, 
$\pi^{\mu\nu}$ is the tensor dissipation or shear stress tensor
and 
$\Delta^{\mu\nu}=g^{\mu\nu}+u^{\mu}u^{\nu}$ is the projection operator,
such that $\Delta^{\mu\nu}u_\nu=0$. 
In Eckart frame, $n^{\mu}=0$ and $q^{\mu}=h^{\mu}$. The relaxation equation for additional 
scalar soft mode $\phi$ is introduced as~\cite{Stephanov:2017ghc}:
\begin{equation}
D\phi=-F_{\phi}+A_{\phi}\theta\,,
\end{equation}
where, $\theta=\partial_{\mu}u^{\mu}$ and $D=u^{\mu}\partial_{\mu}$. Forms of $F_{\phi}$ and $A_{\phi}$ can be obtained 
by imposing second law of thermodynamics which states: 
\beqa
\partial_{\mu}s^{\mu}\geq 0 \,,
\eeqa
where, $s^{\mu}$ is the entropy four current and is defined as
\beqa
s^{\mu}=s u^{\mu}+\Delta s^{\mu}\,.
\eeqa
In hydro+ formalism,  the partial equilibrium entropy gets 
separate contribution from the slow modes represented by, $\phi$. The change in entropy density 
due to the introduction the scalar field, $\phi$ is  given by,
\begin{equation}
ds_{+}=\beta_{+}d\epsilon-\alpha_{+} dn-\pi d\phi\,,
\end{equation}
where, $\pi$ is the `energy cost' for addition of $\phi$ modes in the system 
can be called the corresponding chemical potential. Here  $\alpha_{+}=\mu_{+}/T$, where, $T$ is the temperature 
$\mu_{+}$ is the chemical potential. 

The role of non-hydrodynamic mode on the extensivity condition was ignored in 
earlier calculations.  
In the present work it is found that, this condition is important 
for determining the form of $A_{\phi}$. From here onward, we drop the subscript ``$+$" and
write the following relations in presence of $\phi$ as:
\beqa
dP&=&sdT+n d\mu+\phi d\pi\,,\\
sT&=&\epsilon+P-\mu n-\pi \phi\,.
\label{eq11}
\eeqa
The equation for the soft mode is already of relaxation type in its form, therefore,  
the coupling of slowly evolving soft modes at first order is adequate to maintain causality.
The coupling with the soft mode at first order can be achieved as follows:
\begin{equation}
\partial_\mu s^{\mu}=(F_{\phi}-b  \partial_{\mu} q^{\mu})\pi -q^{\mu}\partial_{\mu}(\beta+b \pi)-\beta (\partial_{\mu} u_{\nu})\Delta T^{\mu\nu}+\partial_{\mu}(\Delta s^{\mu}+\beta q^{\mu}+b \pi q^{\mu})+S_n \theta \,,
\end{equation}
where $b$ represents the coupling strength and 
\begin{equation}
S_n=s-\beta(\epsilon+P)+\mu \beta n+\pi A_{\phi}\,.
\label{eq13}
\end{equation}
In the Eckart frame the soft mode couples with the heat flux. 
$\partial_{\mu}s^{\mu}\geq 0$ can be satisfied, with $S_n=0$ and redefinition of $\pi$, 
$q^{\mu}$, $\Delta T^{\mu\nu}$ and $s^{\mu}$, therefore, we have
\beqa
\Delta s^{\mu}=-\beta q^{\mu} -b \pi q^{\mu}\,,
\eeqa 
and
\beqa
q^{\mu}&=&-\kappa T[D u^{\mu}-\frac{1}{\beta}\Delta^{\mu\nu}\partial_{\nu}(\beta+b \pi)]\,,\\
F_{\phi}&=&\gamma\pi -b \partial_{\mu}[ \kappa T D u^{\mu}-\frac{\kappa}{\beta}\Delta^{\mu\nu}\partial_{\nu}(\beta+b \pi)]\,,
\eeqa
where the proportionality constants $\kappa\ge 0$, $\gamma\ge 0$ are respectively 
thermal conductivity and relaxation rate of slow modes. 
Comparison of Eqs.\eqref{eq11} 
and \eqref{eq13} gives $A_{\phi}=\phi$ for $S_n=0$. 
Therefore, the modified equations for fluxes in the second order IS theory with the
coupling to soft modes read as follows:
\begin{eqnarray}
&&\Pi=-\frac{1}{3}\zeta\Big[\partial_{\mu}u^{\mu}+\beta_0 D\Pi-\tilde{\alpha}_0\partial_{\mu}q^{\mu}\Big]\,,\nonumber\\
&& q^{\mu}=-\kappa T \Delta^{\mu\nu}\Big[\beta\partial_{\nu} (T+b \pi)+Du_{\nu}+\beta_1 D q_{\nu}-\tilde{\alpha}_0\partial_{\nu}\Pi -\tilde{\alpha}_1\partial_{\lambda} \pi^{\lambda}_{\nu}\Big]\,,\nonumber\\
&&\pi^{\mu\nu}=-2\eta\Big[\Delta^{\mu\nu\rho\lambda}\partial_{\rho}u_{\lambda}+\beta_2 D \pi^{\mu\nu}-\tilde{\alpha}_1\Delta^{\mu\nu\rho\lambda}\partial_{\rho}q_{\lambda}\Big]\,,
\end{eqnarray}
with constants of proportionality $\eta\ge0$, $\zeta\ge0$ where $\eta$ and $\zeta$ are the shear and bulk 
viscous coefficients respectively.
The quantities, $\tilde{\alpha_0}$ and $\tilde{\alpha_1}$ are coupling coefficients, 
$\beta _0,\tilde{\beta_1}, \beta_2$ are relaxation coefficients. 
The relations of $\beta _0,\tilde{\beta_1}$ and $\beta_2$ to the relaxation time scales are 
given by ~\cite{Muronga:2003ta,Muronga:2001zk}:
\begin{equation}
\tau_{\Pi}=\zeta \beta_0, \,\,\,\,\tau_q=\kappa T\beta_1,\,\,\,\, \tau _{\pi}=2\eta \beta_2\,,
\label{eq12}
\end{equation}
The coupling coefficients, which couple to heat flux, and bulk pressure $(l_{q\Pi}, l_{\Pi q})$, the heat flux, and shear tensor $(l_{q\pi}, l_{\pi q})$, are related to the relaxation lengths by the following relations,
\begin{equation}
l_{\Pi q}=\zeta \alpha_0,\,\,\,\, l_{q\Pi}=\kappa T \alpha _0, \,\,\,\,l_{q\pi}=\kappa T\alpha_1,\,\,\,\, 
l_{\pi q}=2\eta \alpha_1~.
\label{eq19} 
\end{equation}  
The  expressions for relaxation and coupling coefficients are taken from Ref.~\cite{Hasan2}.

\subsection{Linearized equations and the dynamic structure factor}
Next we linearize the hydrodynamic equations for small deviations from equilibrium 
field quantities. 
We assume $\mathcal{Q}=\mathcal{Q}_{0}+\mathcal{\delta Q}$, 
where $\mathcal{Q}$, $\mathcal{Q}_{0}$ and $\delta\mathcal{Q}$ 
represents hydrodynamic variables, its average value and the fluctuation respectively.  
$\mathcal{Q}_{0}=0$ is its average value for dissipative degrees 
of freedom and $u_0^{\mu}=(-1,\, 0,\, 0,\, 0)$ and $\delta u^{\mu}=(0,\,\delta \vec{u})$. 

In the linearized domain the equations of motion read as:  
\begin{subequations}
\begin{eqnarray}
0&=&-\frac{\partial \delta \epsilon}{\partial t}-(\epsilon_0+P_0)+\vec{\nabla}\cdot \delta \vec{u}-\vec{\nabla} \cdot \delta \vec{ q}\,, \\
 0&=&  -(\epsilon_0+P_0)\frac{\partial}{\partial t} \delta u^{i}-\partial^{i}(\delta P+\delta \Pi)+\frac{\partial}{\partial t} \delta q^{i}-\partial_{j}\Pi^{ij}\,,\\
 0&=&  -\frac{\partial}{\partial t} \delta n-n_0\vec{\nabla}\cdot \delta \vec{u}\,,\\
 0&=&  \delta \Pi+\frac{1}{3}\zeta [\vec{\nabla}\cdot \delta \vec{u}+\beta_0\frac{\partial}{\partial t}\delta \Pi-\tilde{\alpha_0}\vec{\nabla}\cdot \delta \vec{q}]\,,\\
0&=&   \delta q^{i}+\kappa T_0\nabla^{i}\delta T+\kappa T_0 \frac{\partial}{\partial t} \delta u^{i}+\kappa T_0 \beta_1 
   \frac{\partial}{\partial t}\delta q^{i}-\kappa T_0 \tilde{\alpha}_0\nabla^{i}\delta \Pi -\kappa T_0\tilde{\alpha}_1\nabla_{j}\pi^{ij}\,,\\
0&=&   \delta \pi^{ij}+2\eta\delta^{ijlm}(\partial_{l}\delta u_{m}-\tilde{\alpha}_1\partial_{l}\delta q_{m})+2\eta \beta_2\frac{\partial}{\partial t} \delta \pi^{ij}\,,\\
 0&=& - \frac{\partial}{\partial t}\delta \phi-(\gamma+T_0^2\frac{K_{q\pi}^2}{\kappa}\nabla^2)C_{\phi\pi}\delta \phi -
   \big[\gamma+(T_0^2\frac{K_{q\pi}^2}{\kappa}-\frac{K_{q\pi}}{C_{T \pi}})\nabla^2\big] C_{T \pi}\delta T\nn\\
   &&-(\gamma+T_0^2\frac{K_{q\pi}}{\kappa}\nabla^2)C_{\pi n}\delta n-T_0K_{q\pi}\frac{\partial}{\partial t}(\nabla_{i}\delta u^{i})+\tilde{\phi}\nabla_{i}\delta u^{i},
\end{eqnarray}
\end{subequations}
where, $K_{q\pi}=b \kappa$ and
\beqa
C_{ A\pi}=\frac{\partial \pi}{\partial A}, \,\,\,\, \text{with}\,\,\, A\equiv(T,n,\phi)\,.
\eeqa
By applying Fourier-Laplace transformation {\textit{i.e}}, 
$\displaystyle{\lim_{x \to \infty}} \int d^3x\int_{0}^{\infty} \exp[{(z-\epsilon)t-i\vec{k}\cdot\vec{x}}]$ 
from the left on the above equation and performing the integration, then, 
replacing $z$ by $-i\omega$ we get a set of linear equations  
in $\omega-k$ space given in the Appendix~\ref{appendixA}.

The set of equations given in Eqs. \eqref{eq23}-\eqref{eq33} can be written in matrix form as
\beqa
\mathds{M}\delta\mathcal{Q}=\mathcal{A}\,,
\eeqa
where, $\mathds{M}$, is an $11 \times 11$ matrix, arranged row wise of coefficients of $\delta n,\delta T,\delta u_{||}, \delta u_{\perp},\delta \Pi, \delta q_{||}$,\\$\delta q_{\perp},\delta \pi_{||\,||},\delta \pi_{||\,\perp},\delta \pi_{\perp\,\perp}$, and $\delta \phi$ respectively
The set of linear equations can be solved as
\begin{equation}
\delta\mathcal{Q}=\mathds{M}^{-1}\mathcal{A}\,,
\label{delQ}
\end{equation}
In the present work we are interested in evaluating the two point 
correlation of density fluctuation. 
The solution of the set of equations represented by Eq.~\eqref{delQ} leads to the 
following expression for density fluctuation ($\delta n$),
\beqa
\delta n(\vec{k},\omega)=\mathds{M}^{-1}_{11}\big[-\big(\frac{\pd{\epsilon}}{\pd n}\big) \delta n (\vec{k}, 0)-\big(\frac{\pd{\epsilon}}{\pd T}\big) \delta T (\vec{k}, 0)-\big(\frac{\pd{\epsilon}}{\pd \phi}\big) \delta \phi (\vec{k}, 0)\Big]-\mathds{M}^{-1}_{14}\big[-\delta n(\vec{k}, 0)\Big].
\eeqa
Now we define the correlation of density fluctuations, as $\mathcal{S^\prime}_{nn}(\vec{k},\omega)$ by the 
following expression: 
\beqa
\mathcal{S^\prime}_{nn}(\vec{k},\omega)=\Big< \delta n(\vec{k},\omega)\delta n(\vec{k},0)\Big>\,.
\label{eq38}
\eeqa
Since, the correlation between two independent thermodynamic variables, say, $\mathcal{Q}_i$ and $\mathcal{Q}_j$  vanishes i.e
\beqa
\Big< \delta \mathcal{Q}_{i}(\vec{k},\omega)\delta \mathcal{Q}_{j}(\vec{k},0)\Big>=0, \,\,\,\, i\neq j \,.
\label{eq39}
\eeqa
Therefore, $\mathcal{S}^{'}_{nn}$ can be written as
\beqa
\mathcal{S}^{'}_{nn}(\vec{k}, \omega)=-\Big[\big(\frac{\pd{\epsilon}}{\pd n}\big) \mathds{M}^{-1}_{11}-\mathds{M}^{-1}_{14}\Big] \Big< \delta n(\vec{k}, 0)\delta n(\vec{k},0)\Big>\,,
\eeqa
The final expression for the $\Snn$ can be obtained as:
\beqa
\mathcal{S}_{nn}(\vec{k}, \omega)=\frac{\mathcal{S}^{'}_{nn}(\vec{k}, \omega)}{\Big< \delta n(\vec{k}, 0)\delta n(\vec{k},0)\Big>}=-\Big[\big(\frac{\pd{\epsilon}}{\pd n}\big) \mathds{M}^{-1}_{11}-\mathds{M}^{-1}_{14}\Big]\,.
\eeqa
This is the spectral correlation  in density fluctuation with the inclusion 
of the extra degree of freedom, $\phi$.

\section{Results and discussion}
\label{sec3}
\begin{figure}[h]
\centering
\includegraphics[width=8.1cm]{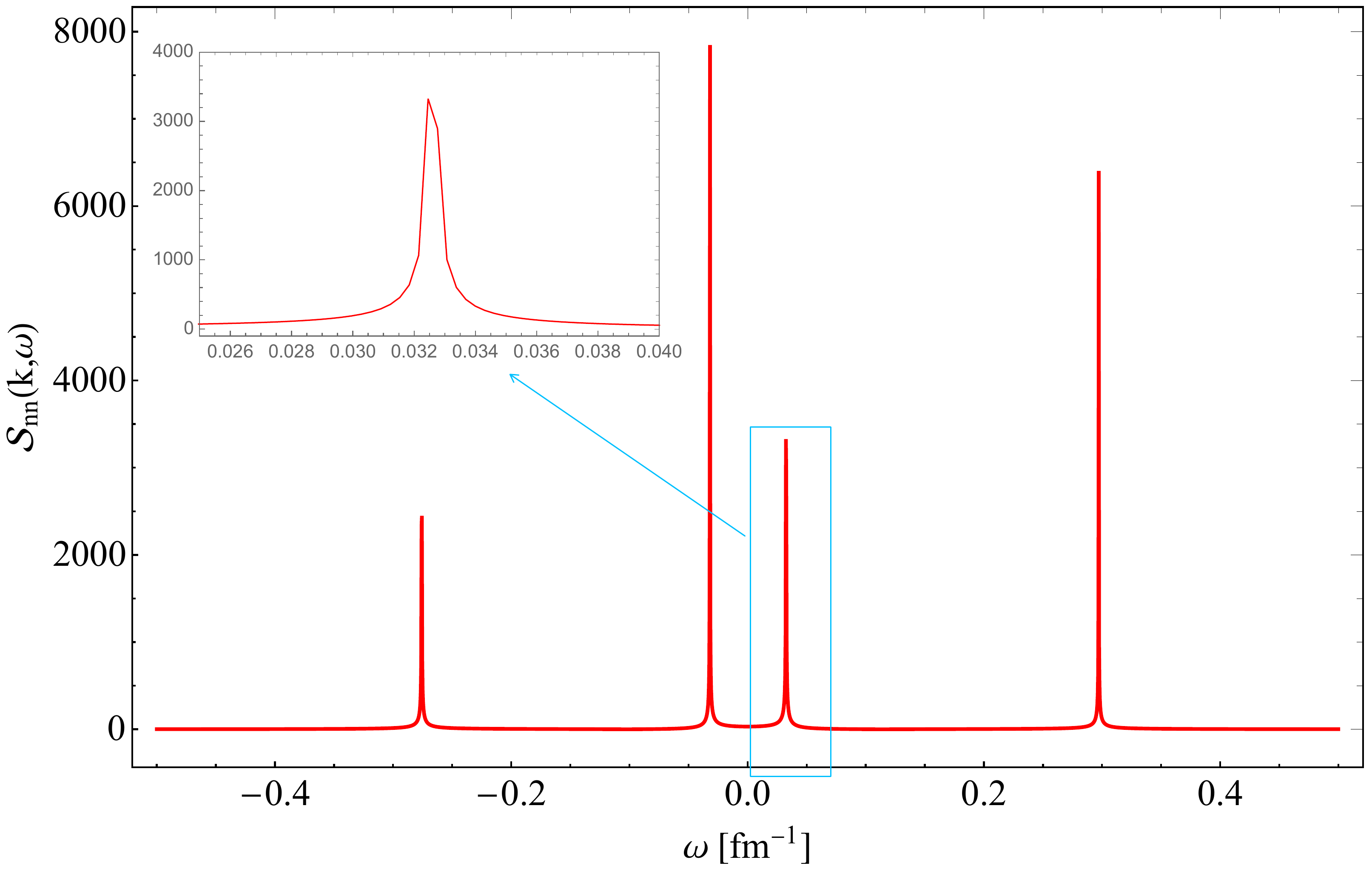}	
\includegraphics[width=8.1cm]{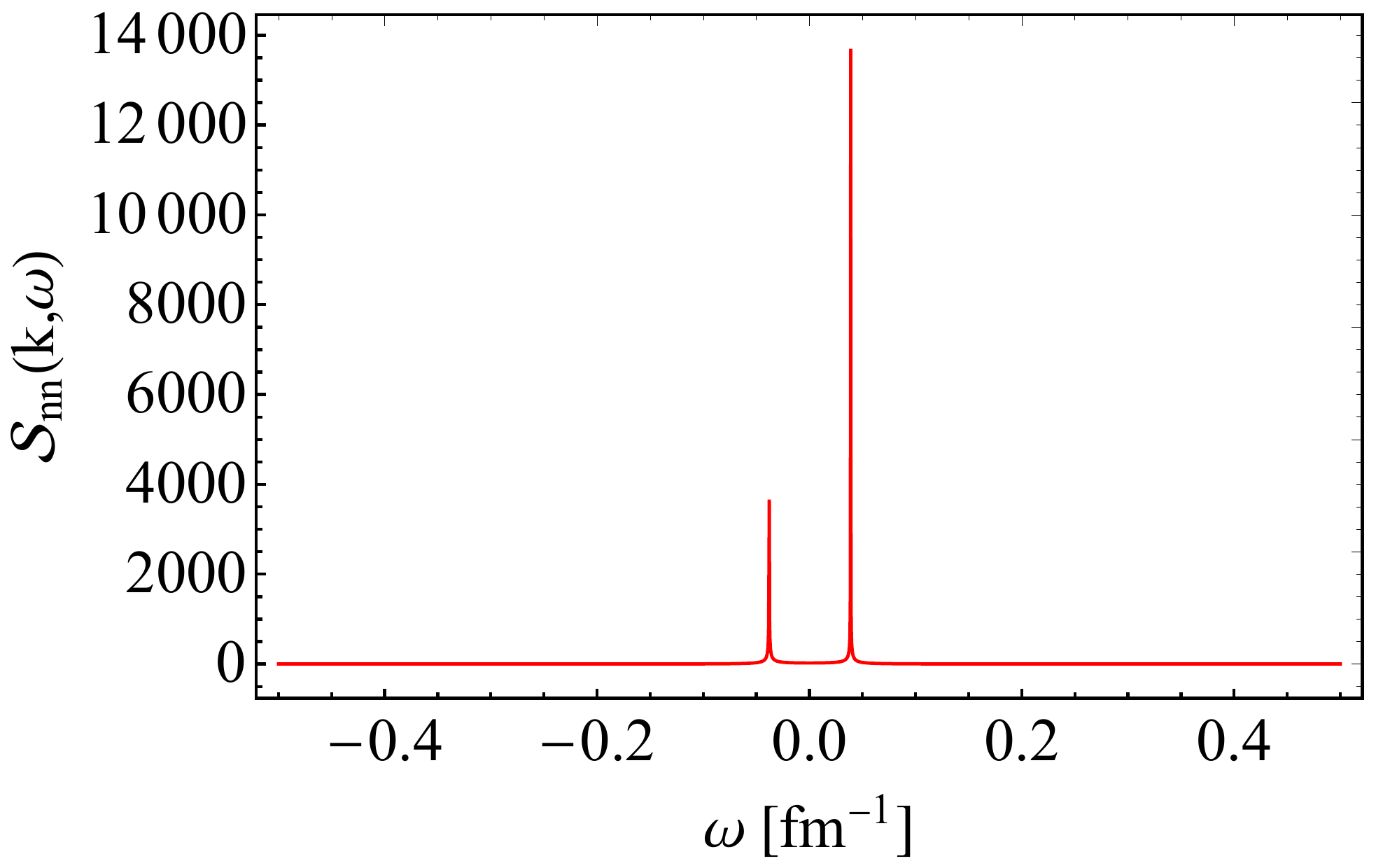}	
\caption{(Color online)  (a) Variation of $\Snn$ with 
$\omega$ for  $k=0.1\, fm^{-1}$ and $\eta/s= \zeta/s =\kappa T/s=1/4\pi$, 
when the system is away from CEP ($r=0.2$). The peaks are appearing 
as very narrow due to the  scale chosen along x-axis, change in x-axis scale
makes the width visible (see the inset for the R-peak). 
(b) The system is close to the CEP ($r=0.01$). 
The results are obtained with the EoS containing the CEP and the transport coefficients,
thermodynamic response function and the relaxation coefficients 
are estimated from the scaling behaviour.}
\label{fig1}
\end{figure}

The aim of this work is to find out the spectral structure  of density fluctuation
within the scope of relativistic causal hydrodynamics with its validity extended 
near the CEP by introducing a scalar field as discussed above.  
The effect of the CEP is taken into consideration
through an Equation of State (EoS), containing the critical point. 
The EoS is constructed from the  universality hypothesis which suggests that the 
CEP of the QCD belongs to same universality class as the 3D 
Ising model. We do not repeat the discussion on the construction of the EoS here but
refer to the appropriate literature~\cite{Parotto, Nonaka,Hasan1,Hasan2} for details. 
The behaviour of various transport coefficients and response functions 
play crucial roles to determine the structure factor in the presence of the CEP.
The scaling behaviour of various transport coefficients and thermodynamic functions near the CEP
have been taken~\cite{KapustaChi,Guida,Rajagopal:1992qz} into account by using the
following relations:
\beqa
&&\kappa_{T}=\kappa_{T}^{0}|r|^{-\gamma^\prime}, C_{V}=C_{0}|r|^{-\alpha}, 
C_{P}=\frac{\kappa_{0}T_{0}}{n_{0}}\Big(\frac{\pd P}{\pd T}\Big)^{2}_{n}|r|^{-\gamma^\prime},\nn\\
&& c^{2}_{s}=\frac{T_{0}}{n_{0}h_{0}C_{0}}\Big(\frac{\pd P}{\pd T}\Big)^{2}_{n}|r|^{\alpha},
\alpha_{p}=\kappa_{0}\Big(\frac{\pd P}{\pd T}\Big)_{n}|r|^{-\gamma^\prime} \nn\\
&& \eta=\eta_{0}|r|^{1+a_{\kappa}/2-\gamma^\prime}, \zeta= \zeta_{0}|r|^{-\alpha_{\zeta}},
\kappa=\kappa_{0}|r|^{-a_{\kappa}}\,,
\label{eq36}
\eeqa
where $r=(T-T_c)/T_c$ is the reduced temperature and 
$\alpha,\, \gamma^\prime,\, a_{\zeta},\, a_{\kappa}$ are the critical exponents.
Here $\alpha=0.11$ and $\gamma^\prime= 1.2$.
We take the values $a_{\zeta}=\nu d-\alpha=1.78$
(here, $d=3, \nu=0.63$), and $a_{\kappa}=0.63$. The partial derivative,
$\left(\frac{\pd P}{\pd T}\right)_n$ has been evaluated by using the EoS
for consistency.
The critical behaviours of the second-order coupling and relaxation coefficients 
of the IS hydrodynamics have also been taken into account (see~\cite{Hasan2} for details).
\begin{figure}[h]
\centering
\includegraphics[width=10.1cm]{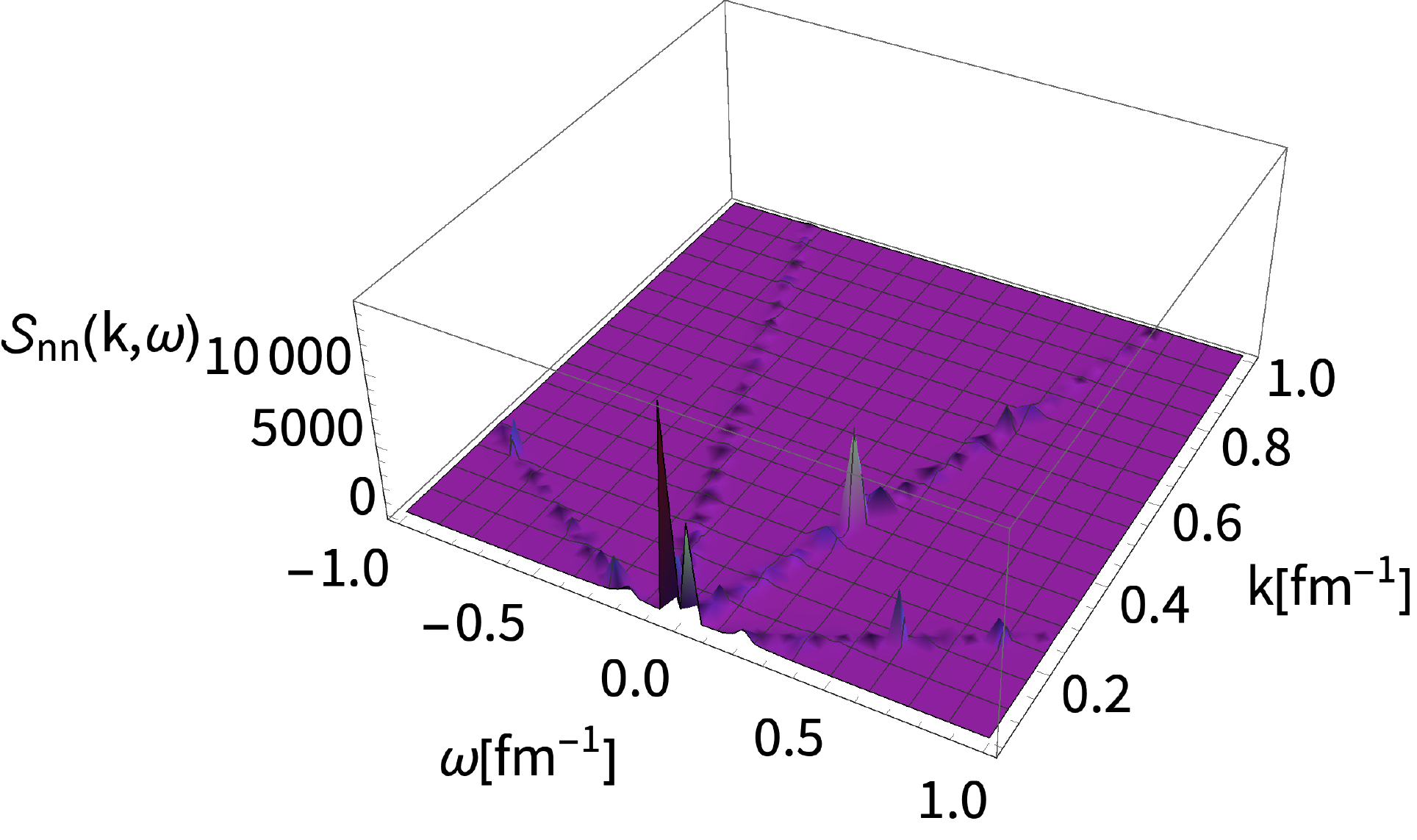}	
\caption{(Color online) Variation of $\Snn$ with $\omega$ and $k$ for $r=0.2$, 
i.e when the system is away from CEP.}
\label{fig2}
\end{figure}
\begin{figure}[h]
\centering
\includegraphics[width=10.1cm]{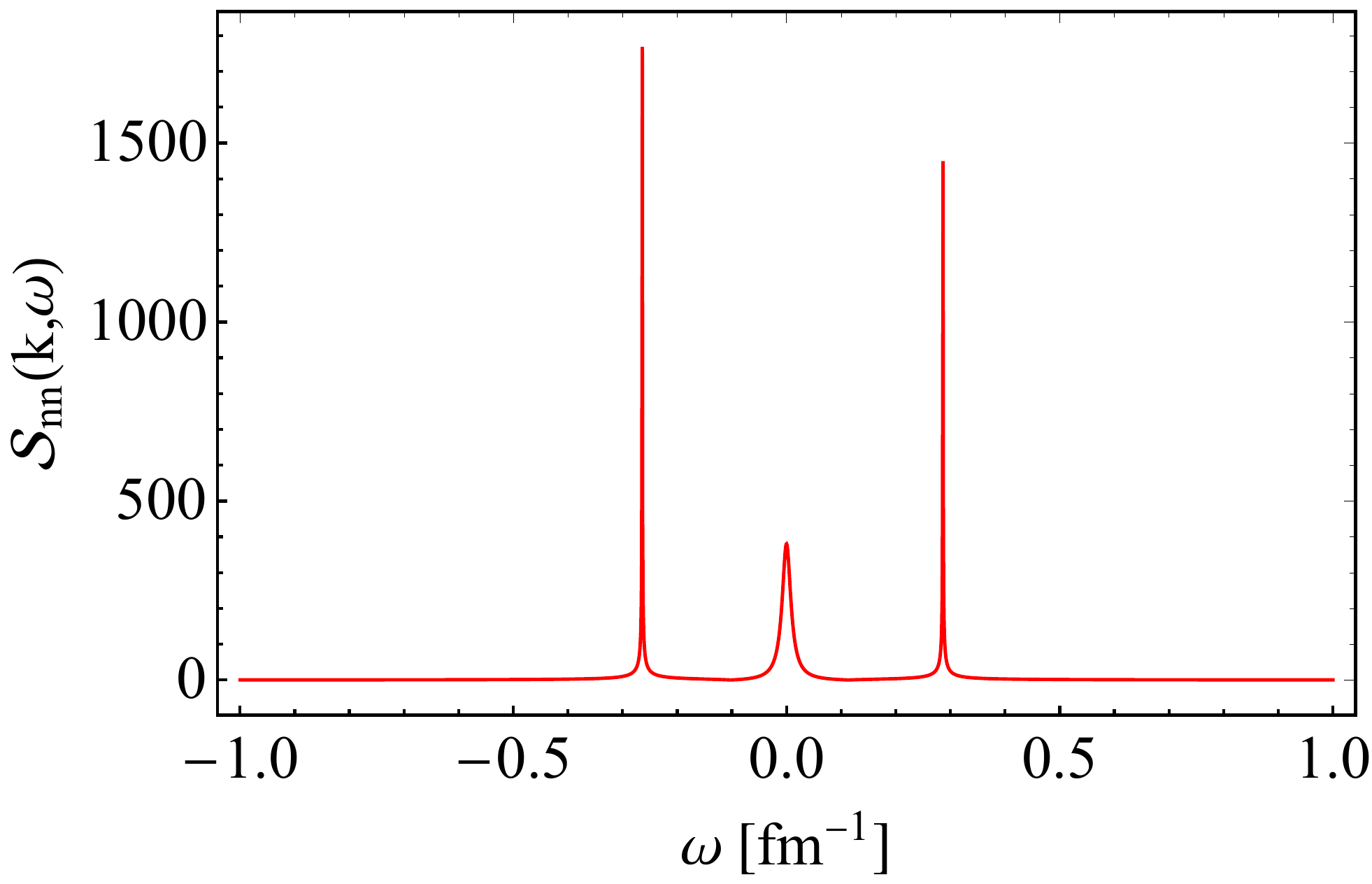}	
\caption{(Color online) Variation of $\Snn$ with $\omega$ for $k=0.1\, fm^{-1}$ and  
$r=0.2$, i.e the system is away from CEP for $K_{q\pi}=0.$}
\label{fig3}
\end{figure}

In Fig.\ref{fig1}, the variation of $\Snn$ with $\omega$ is shown for $k=0.1$ fm$^{-1}$,
when the system is away from the CEP (left panel). In absence of $\phi$, the $\Snn$ 
has three peaks. These are the  R-peak  located at $\omega=0$ and the B-peaks 
located symmetrically on either sides of the R-peak.  The $\Snn$  admits four peaks 
in the presence of $\phi$ which is introduced to extend the validity of hydrodynamics. 
It is interesting to note that there is no elastic peak (R-peak) at $\omega=0$. 
Two away side peaks are identified as the Brillouin peaks (B-peaks). The Stokes 
component (left side) and the anti-Stokes component 
(right side) are located asymmetrically on either side
of the origin with unequal magnitudes. The other peak is 
due to the coupling of $\phi$ with the heat flux. 
This profile of  $\Snn$ is different from
the one evaluated  earlier~\cite{Hasan2}, in absence of $\phi$, 
where the Stokes and anti-Stokes components of equal magnitude are symmetrically 
located about $\omega=0$.  
The asymmetry in the B-peaks may arise due to the local inhomogeneity 
present in the system. The B-peaks arise from propagating sound modes 
associated with pressure fluctuations at constant entropy. 
In condensed matter physics, the asymmetry of the 
B-peaks are understood from the fact that two sound modes with different
$\omega$ values, $-c_sk$ and $+c_sk$
originate from  different temperature zone's~\cite{Rayeligh_Benard,Zarate}. 
The asymmetry present in Rayleigh 
component  is exactly opposite to the Brillouin components. 
In the vicinity of the CEP (right panel) the $\Snn$ 
shows two peaks as the B-peaks disappear due to the absorption
of sound at the CEP. 

In Fig.~\ref{fig2} the structure factor in $\omega-k$ plane has been plotted
when the system is away from CEP. The structure of $\Snn$ with non-zero $\phi$ 
is quite different from the case when $\phi=0$. In case of $\phi=0$ 
there is a $R$-peak and two $B-$ peaks
at $(\omega=0)$ and $(\omega=pm c_sk)$ respectively with
diminishing  peak values as $k$ increases. However, with $\phi\neq 0$ 
peaks appear at non-zero $k$ values too due to the coupling of
$\phi$ with other hydrodynamic fields.
The symmetric structure of $\Snn$ with R and B peaks can be
reproduced when the coupling of $\pi$ (the chemical potential
corresponding to the variable $\phi$) with flux $q$ is set to zero
(Fig.~\ref{fig3}). 

\begin{figure}[h]
\centering
\includegraphics[width=10.1cm]{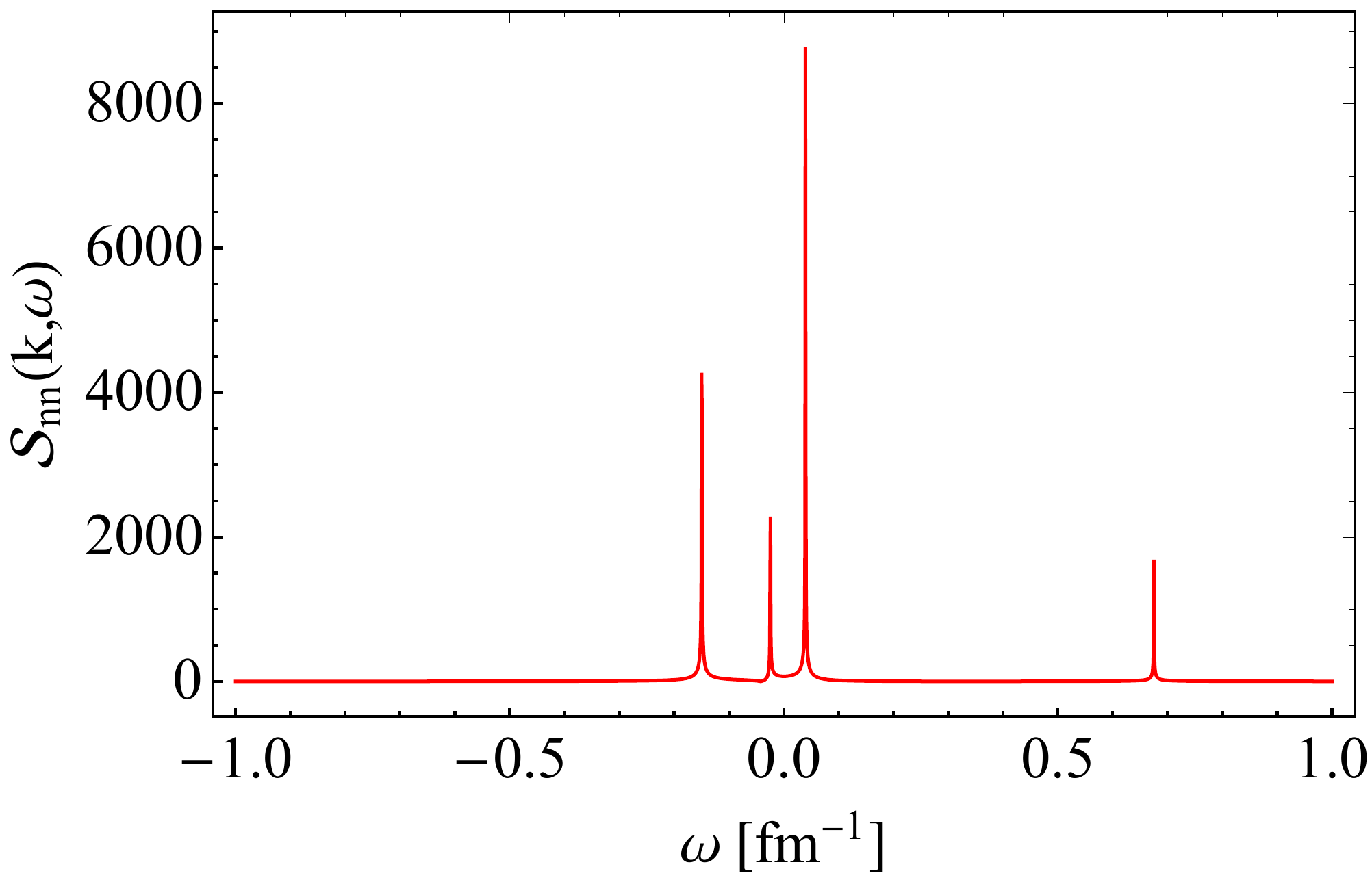}	
\caption{(Color online) Variation of $\Snn$ with $\omega$ for $k=0.1\, fm^{-1}$ 
and $r=0.2$, i.e the system is away from CEP  but with higher value of 
$(\frac{\pd P}{\pd \phi})$. The asymmetry in the B-peaks is clearly visible.}
\label{fig4}
\end{figure}
The asymmetry in $\Snn$ with respect to $\omega$ 
increases with  increase in $\big(\partial P/\partial\phi\big)$.
This is distinctly visible in Fig.~\ref{fig4} in comparison
to results displayed in Fig.~\ref{fig1}. The locations of
the four peaks in $\omega$ are obtained from the dispersion relation
provided in the Appendix~\ref{appendixB}. 

\begin{figure}[h]
\centering
\includegraphics[width=10.1cm]{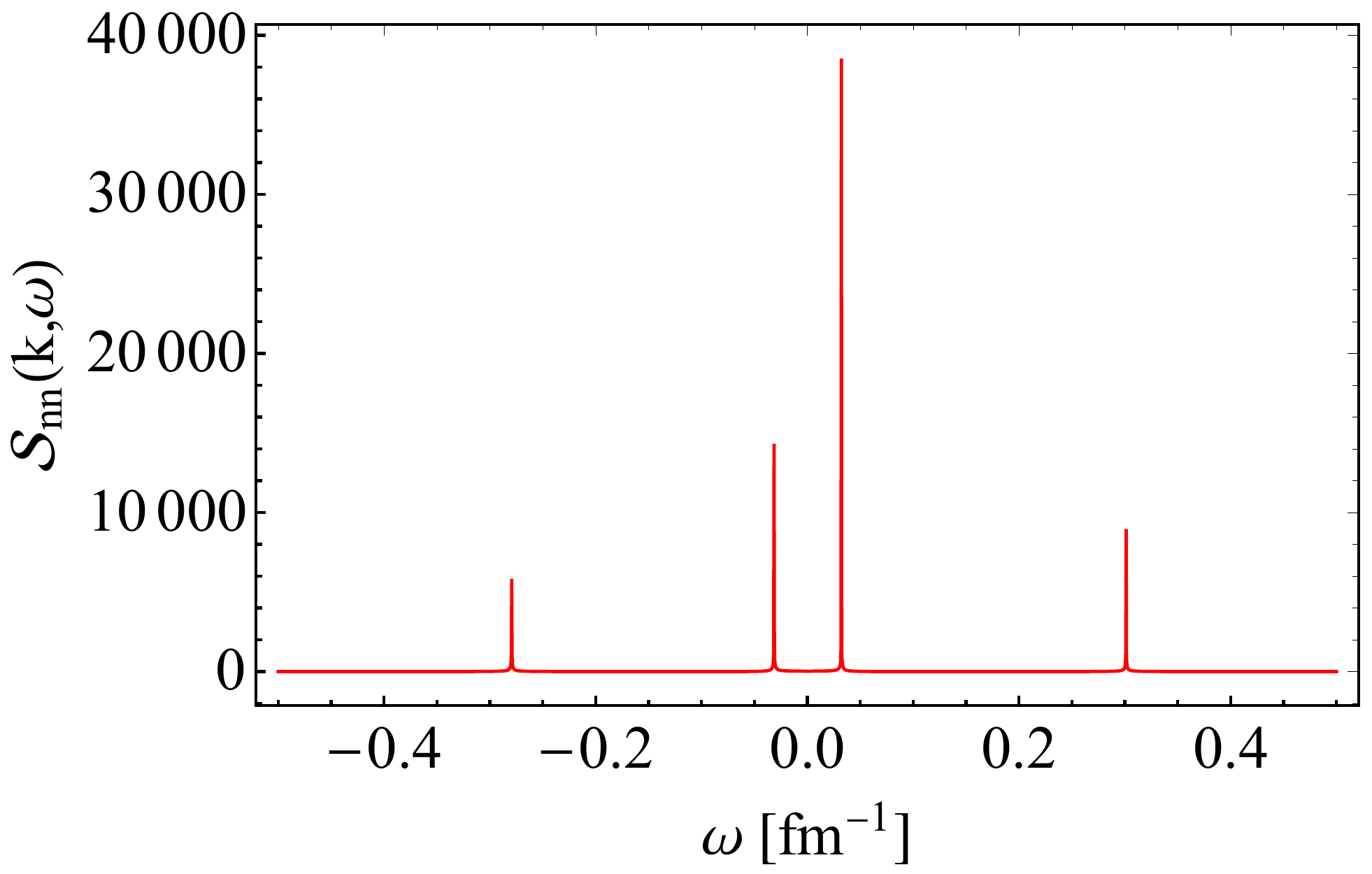}	
\caption{(Color online) Variation of $\Snn$ with $\omega$ for $k=0.1\, fm^{-1}$ 
and $r=0.2$, i.e the system is away from CEP with increased $\phi$ modes 
in the system which increases the thermal fluctuation.}
\label{fig5}
\end{figure}
Fig.~\ref{fig5} displays the structure factor with increased $\phi$ 
mode. The height of the peaks get enhanced significantly  due to
the enhancement in $\phi$ value.

\begin{figure}[h]
\centering
\includegraphics[width=10.1cm]{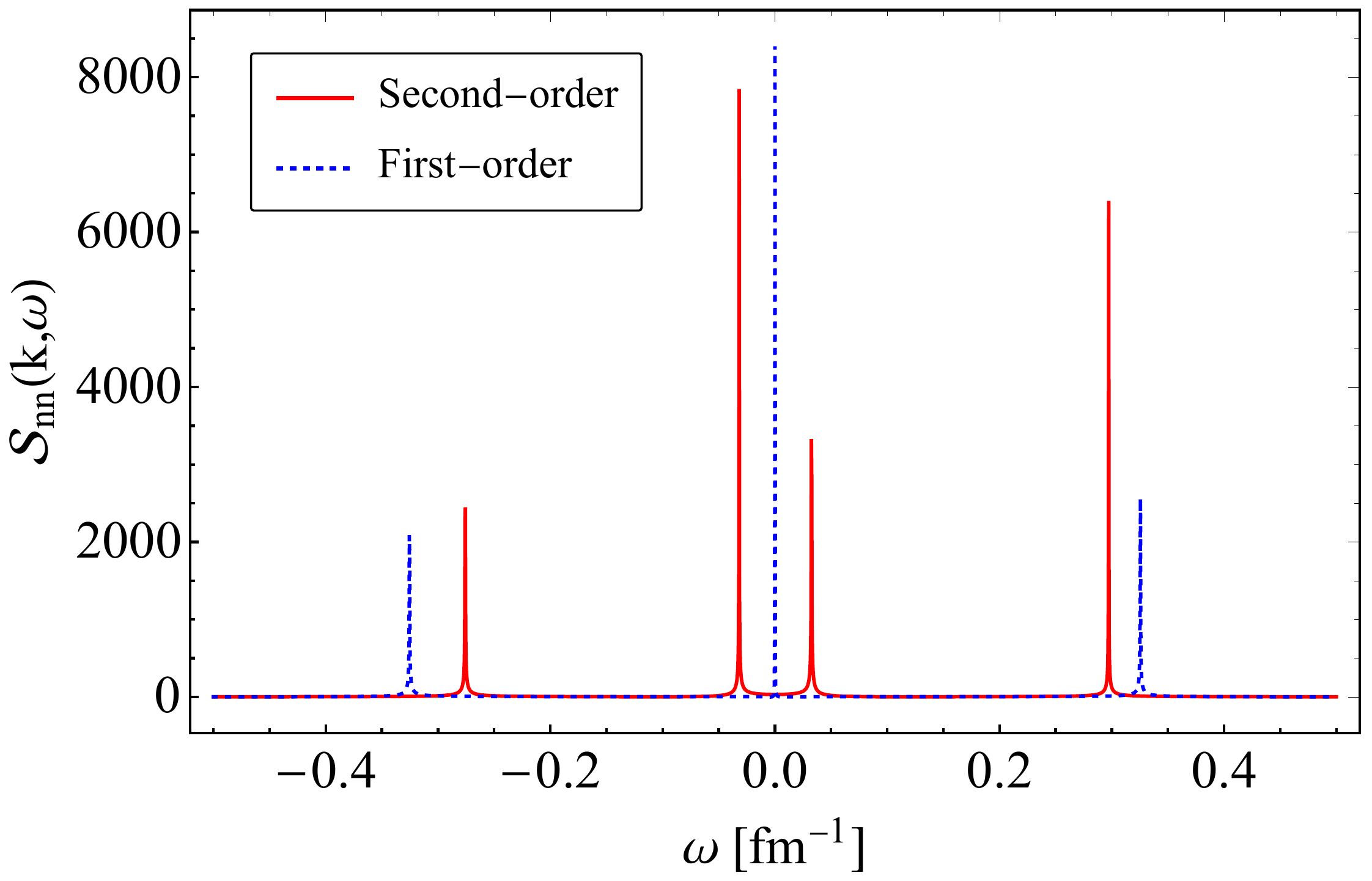}	
\caption{(Color online) Variation of $\Snn$ with $\omega$ for
$k=0.1\, fm^{-1}$ and $r=0.2$. The results for second order
and first order hydrodynamics are  compared here.}
\label{fig6}
\end{figure}
In Fig.~\ref{fig6} the structure factor for second-order hydrodynamics
has been compared with first-order hydrodynamics (relativistic Navier-Stokes).
Interestingly the structure factor for first-order hydrodynamics admits 
a R-peak at the origin and two symmetric B-peaks located on the opposite
sides of the R-peak. The effects of $\phi$ seems to be inconsequential in the
first-order theory because of the vanishing of various 
coupling and relaxation coefficients.

\begin{figure}[h]
\centering
\includegraphics[width=10.1cm]{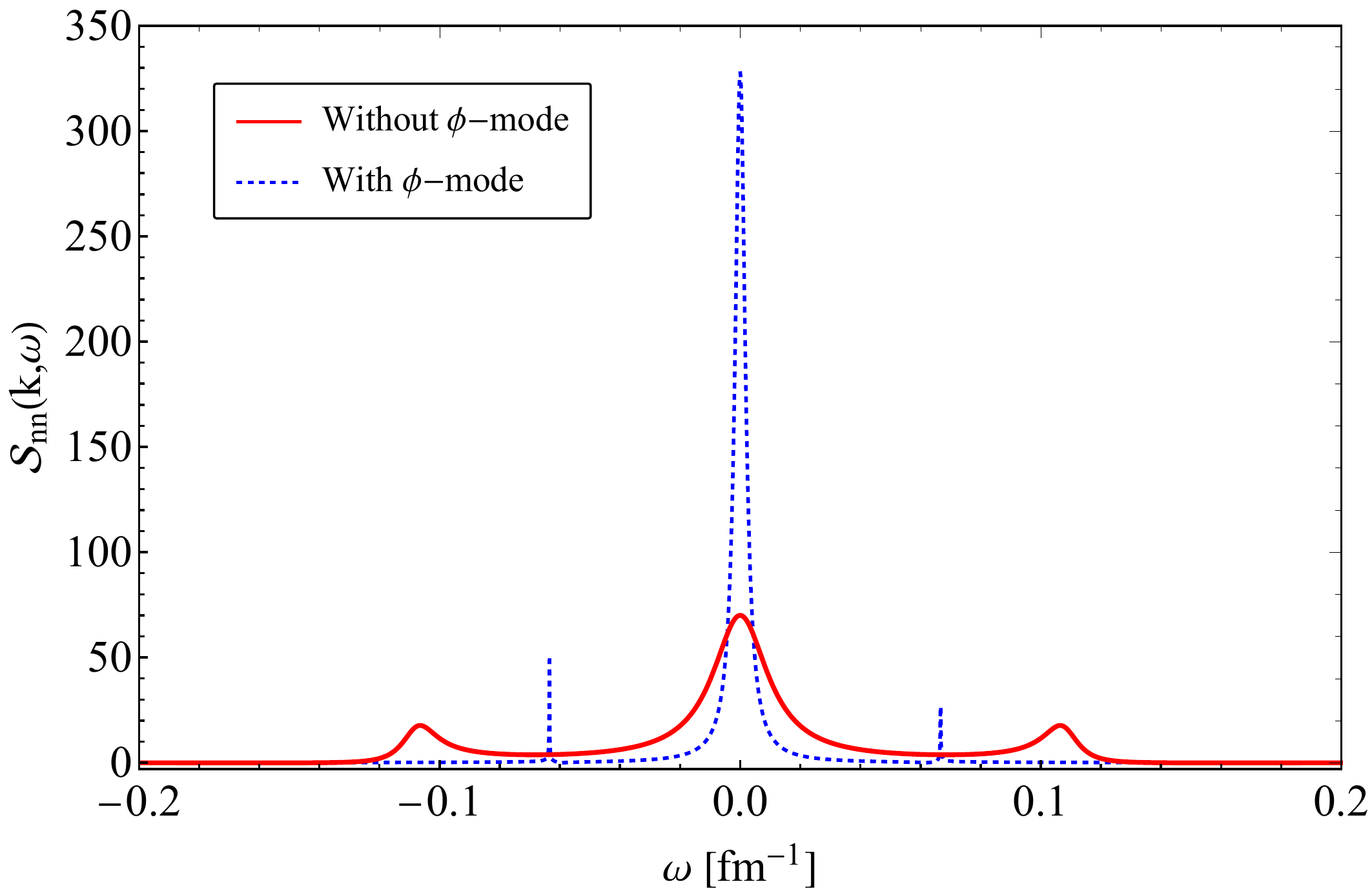}	
\caption{(Color online) Variation of $\Snn$ with $\omega$ for
$k=0.1\, fm^{-1}$ and $r=0.2$ with (dotted line) and without (solid)
$\phi$ mode, when the longitudinal modes are only considered.
}
\label{fig7}
\end{figure}
The $\omega$ dependence of $\Snn$ derived from the longitudinal dispersion relation with and without $\phi$ has
been depicted in Fig.~\ref{fig7}. We find that the width of the B-peaks are  small due to the
introduction of the $\phi$ field (blue dashed line) and closer to the R-peak compared to
the case when $\phi=0$. The rate of decay of the thermal fluctuation  reflected
through the width of the R-peak becomes smaller  due to the introduction of the 
field $\phi$, that is,  the decay of the fluctuation becomes slower
in the presence of $\phi$.  
The comparison of this result with that displayed in Fig.~\ref{fig1} indicate that the extra 
peak appeared in Fig.~\ref{fig1} is due to the coupling of the transverse modes with $\phi$.
It is interesting to note that the presence of slow modes can cause the extra peak in the dynamic structure 
factor in the presence of the transverse modes in the causal theory of hydrodynamics.

\section{Summary and discussions}
\label{sec4}
It has been shown that the validity of second-order 
hydrodynamics can be extended near the CEP by introducing an extra degree 
of freedom~\cite{Stephanov:2017ghc}. Here we have derived the equation of slow modes for the situation when 
the extensive nature of thermodynamics is not altered due to introduction of slow modes. 
We find that the extensivity condition puts extra constraint on the coupling of scalar slow modes 
to the velocity four divergence. The role of this extra out-of-equilibrium mode  on the spectral 
properties of dynamical density fluctuations near the QCD critical point has been investigated. 
The dynamical spectral structure  ($\Snn$) in presence of the out-of equilibrium 
modes admits four Lorentzian peaks. 
Whereas the dynamic structure factor without out-of-equilibrium modes admits
three Lorentzians peak structure. We find that the asymmetric peaks 
originate due to  coupling of the out-of-equilibrium modes with the hydrodynamic modes. 
It is also shown that the first-order hydrodynamics 
is unaffected by the extra variable introduced to broaden the 
scope of hydrodynamics. The presence of slow modes can cause the extra peak in the dynamic 
structure factor due to the presence of the transverse modes in the causal theory of hydrodynamics.
These effects of slow modes on the dynamic structure factor may help in the experimental investigation 
of role of slow modes near the $\mathcal{O}(4)$ critical points other than that is expected in heavy 
ion collisions, such as condensed matter systems. Consequently, by virtue of universality class, 
the acquired knowledge on the role of slow modes from experiments on the critical point 
may be useful in estimating the effect of the QCD CEP through modeling the hydrodynamic 
evolution of system formed in heavy ion collisions near the QCD critical point. 
The  field representing the OEM, $\phi$ plays a crucial role. It reduces the width of
the distribution representing the thermal fluctuation which increases the decay time of the
fluctuation. 
\section{Acknowledgement}
\label{sec8}
 GS and MH acknowledge Guruprasad Kadam for fruitful discussions. GS also thanks Hiranmaya Mishra for support during part of this work. MH would also like to thank Department of Higher Education, Govt. of West Bengal, India. 

\appendix
\section{The set linear equations derived for the perturbations in $\omega-k$ space}
\label{appendixA}
The set of equations for the perturbations in hydrodynamic and non-hydrodynamic
($\phi$) fields in $\omega-k$ space are given in the followings. Here 
$\delta\epsilon(\vec{k},\omega)$ is in the $\omega$-space and $\delta\tilde{\epsilon}(\vec{k},t=0)$ is
defined in $t$-space. Similar notation has been used for other fields.

\begin{eqnarray}
i\omega \delta \epsilon(\vec{k}, \omega) + i k \delta q_{||}(\vec{k}, \omega)+i k(\epsilon_0+P_0)\delta u_{||}(\vec{k},\omega)= 
-\delta \tilde{\epsilon}(\vec{k}, t=0),
\label{eq23}
\eeqa

 \beqa
  i\omega (\epsilon_0+P_0)\delta u_{||}(\vec{k},\omega)-i k \delta P(\vec{k},\omega)-i k \delta\Pi(\vec{k},\omega)&+&i \omega \delta q_{||}(\vec{k},\omega)- i k \delta \pi_{||\,||}(\vec{k},\omega)  \nn\\
   &=&(\epsilon_0+P_0)\tilde{\delta u}_{||}(\vec{k},t=0)-\delta q_{||}(\vec{k},\omega),
   \eeqa
   \beqa
   -i \omega (\epsilon_0+P_0)\delta u_{\perp}(\vec{k},\omega)-i\omega \delta q_{\perp}(\vec{k},\omega)-i k \delta \pi_{\perp \,||}(\vec{k}, \omega)=(\epsilon_0+P_0)\delta u_{\perp}(\vec{k},\omega)-\tilde{\delta q}_{\perp}(\vec{k}, t=0),
   \eeqa
   \beqa
   i\omega \delta n(\vec{k},\omega)+i k n_0 \delta u_{||}(\vec{k},\omega)=-\tilde{\delta n}(\vec{k}, t=0),
  \eeqa
  \beqa
   (1+i\omega\frac{1}{3}\zeta \beta_0)\delta \Pi(\vec{k},\omega)+i k\frac{1}{3}\zeta \delta u_{||}(\vec{k},\omega)-i k \frac{1}{3}\zeta \tilde{\alpha}_0 \delta q_{||}(\vec{k},\omega)=-\frac{1}{3}\zeta \beta_0 \tilde{\delta\Pi}(\vec{k}, t=0),
   \eeqa
  \beqa
   &&(1+i \omega \kappa T_0 \beta_1)\delta q_{||}(\vec{k},\omega)+i k \kappa T_0 \delta u_{||}(\vec{k},\omega)+i k\big[\kappa-C_{T \pi} K_{q\pi} T_0^2\big]\delta T(\vec{k},\omega)-i k T_0^2 K_{q\pi}C_{\phi \pi}\delta\phi(\vec{k},\omega)\nonumber\\ 
   &&-i k\kappa T_0\tilde{\alpha}_0\delta \Pi (\vec{k},\omega)-i k T_0^2K_{q\pi}C_{n \pi}\delta n(\vec{k},\omega)-i k \kappa T_0 \tilde{\alpha}_1 \delta \pi_{||\,||}(\vec{k},\omega)\nn\\
  &&\hspace{3.4in}=-\kappa T_0 \tilde{\delta u}_{||}(\vec{k}, t=0)-\beta_1 \kappa T_0 \tilde{\delta q}_{||}(\vec{k}, t=0),
  \eeqa
  \beqa
 (1+i\omega \beta_1 k T_0)\delta q_{\perp}(\vec{k},\omega)+i\kappa T_0(\omega \delta u_{\perp}(\vec{k},\omega) -k  \tilde{\alpha}_1\delta \pi_{\perp ||}(\vec{k},\omega)\big)=-\kappa T_0\big[\delta u_{\perp}(\vec{k}, 0)-
\beta_1\tilde{\delta q}_{\perp}(\vec{k}, t=0)\big]
  \eeqa
  \beqa
  &&(1+2i\omega\eta \beta_2)\delta \pi_{||\,||}(\vec{k},\omega)+i k\frac{4}{3}\eta \delta u_{||}(\vec{k},\omega)-\frac{4}{3}\delta q_{||}(\vec{k},\omega)=-2\eta\beta_2\tilde{\delta\pi}_{||\,||}(\vec{k}, t=0),
  \eeqa
  \beqa
  && (1+2 i\omega \eta \beta_2)\delta \pi_{\perp\, \perp}
  (\vec{k},\omega)-i k \frac{4}{3} \eta \delta u_{||}(\vec{k},\omega)+i k\frac{4}{3}\eta \tilde{\alpha}_1 \delta q_{||}(\vec{k},\omega)=-2\eta \beta_2 \tilde{\delta\pi}_{\perp\, \perp}(\vec{k}, t=0),
  \eeqa
  \beqa
  &&(1+2 i\omega \eta \beta_2)\delta \pi_{\perp\,\perp}(\vec{k},\omega)+i k \eta \delta u_{\perp}(\vec{k},\omega)-i k \delta q_{\perp}(\vec{k},\omega)=-2\eta \beta_2 \tilde{\delta\pi}_{\perp\,||}(\vec{k}, t=0),
  \eeqa
  \beqa
  && (i\omega +C_{\phi \pi}(\gamma-T_0^2\frac{K_{q\pi}}{\kappa}k^2))\delta \phi(\vec{k},\omega)+\{\gamma-K_{q\pi}(T_0^2\frac{K_{q\pi}}{\kappa}-\frac{1}{C_{T \pi}})k^2\}C_{T \pi}\delta T(\vec{k},\omega)\nn\\
  &&+(\gamma-T_0^2\frac{K_{q\pi}^2}{\kappa}k^2)C_{n \pi}\delta n(\vec{k},\omega)-i k(\tilde{\phi}-i\omega T_0K_{q\pi})\delta u_{||}(\vec{k},\omega)\nn\\
  &&\hspace{2.in}=-(1+i k \frac{K_{q\pi}^2}{\kappa}T_0 C_{\phi \pi})\tilde{\delta\phi}(\vec{k}, t=0)-
i k T_0 K_{q\pi}\tilde{\delta u}_{||}(\vec{k}, t=0)\,,
  \label{eq33}
\end{eqnarray}
where, subscripts "$||$" and "$\perp$" stand for projection along and perpendicular to $\vec{k}$ respectively. After expressing $\delta \epsilon$ and $\delta p$ as:
\beqa
 \delta \epsilon&=&\big(\frac{\partial \epsilon}{\partial n}\big)\delta n+\big(\frac{\partial \epsilon}{\partial T}\big)\delta T+\big(\frac{\partial \epsilon}{\partial \phi}\big)\delta \phi,\\
  \delta P&=&\big(\frac{\partial P}{\partial n}\big)\delta n+\big(\frac{\partial P}{\partial T}\big)\delta T+\big(\frac{\partial P}{\partial \phi}\big)\delta \phi\,.
 \eeqa

\section{Speed of sound and the roots of $\omega(k)$}
\label{appendixB}
In this appendix we provide the expressions for the speed of sound 
and the roots of $\omega(k)$ with the inclusion of 
out-of-equilibrium mode $\phi$.
The speed of sound ($c_s$) is given by,
\beqa
c_s^2&=&\Big(\frac{\pd p}{\pd \epsilon}\Big)_{s/n}\nn\\
&=& \frac{sdT+nd\mu+\phi d\pi}{Tds+\mu dn+\pi d\phi}\nn\\
&=&\mathcal{A}/\mathcal{B}\,,
\eeqa
where, $\mathcal{A}$ and $\mathcal{B}$ are given by 
\beqa
\mathcal{A}= {s+n\mathcal{F} +\phi \mathcal{G} }\,,
\eeqa
\beqa
\mathcal{B}&=&T(\frac{\partial s}{\partial T})_{\mu,\pi}  + T\mathcal{F}(\frac{\partial s}{\partial \mu})_{T,\pi} +T\mathcal{G}(\frac{\partial s}{\partial \pi})_{T,\mu}+\mu (\frac{\partial n}{\partial T})_{\mu,\pi}  + \mu \mathcal{F}(\frac{\partial n}{\partial \mu})_{T,\pi}\nn\\
&+&\mu \mathcal{G}(\frac{\partial n}{\partial \pi})_{T,\mu} +T(\frac{\partial s}{\partial T})_{\mu,\pi}  + T\mathcal{F}(\frac{\partial s}{\partial \mu})_{T,\pi}+T\mathcal{G}(\frac{\partial s}{\partial \pi})_{T,\mu} \,.
\eeqa
where,
\beqa
\mathcal{F}&=& \frac{\Big[(\frac{\partial s}{\partial T})- 
\frac{s}{n}(\frac{\pd n}{\pd T})\Big]+(\frac{\partial \pi}{\partial T})\Big[(\frac{\partial s}{\partial \pi})- 
\frac{s}{n}(\frac{\pd n}{\pd \pi})\Big]}{\frac{s}{n}(\frac{\pd n}{\pd \mu})_{T}
-(\frac{\partial s}{\partial \mu})_T}\,,\\
\mathcal{G}&=& \frac{\Big[(\frac{\partial s}{\partial T})- 
\frac{s}{n}(\frac{\pd n}{\pd T})\Big]+(\frac{\partial \mu}{\partial T})\Big[(\frac{\partial s}{\partial \mu})- 
\frac{s}{n}(\frac{\pd n}{\pd \mu})\Big]}{\frac{s}{n}(\frac{\pd n}{\pd \mu})_{T}
-(\frac{\partial s}{\partial \mu})_T}\,.
\eeqa
The four roots of $\omega$ indicating the location of peaks 
in the structure factor obtained from the dispersion relation are
given below.
\beqa
\omega_{1} =\frac{-\sqrt{-(\epsilon_0+P_0)^2}+i \epsilon_0+i P_0}{2 T_0 \chi }+\frac{\eta  k^2 \sqrt{-(\epsilon_0+P_0){}^2}}{(\epsilon_0+P_0){}^2}+\mathcal{O}(k^3)\,,
\eeqa
\beqa
\omega_{2} =\frac{\sqrt{-(\epsilon_0+P_0){}^2}+i \epsilon_0+i P_0}{2 T_0 \chi }-\frac{\eta  k^2 \sqrt{-(\epsilon_0+P_0){}^2}}{(\epsilon_0+P_0){}^2}+\mathcal{O}(k^3)\,,
\eeqa
\beqa
\omega_{3} &=&\frac{i \gamma  (C_{\phi \pi } \epsilon_T-C_{{T\pi }} \epsilon_{\phi })}{3 \epsilon_T}+\frac{ik^2}{9 \epsilon_T (\epsilon_0+P_0)} \Big[3 T_0^2 \bar{\phi } C_{\phi \pi } \epsilon_T \kappa _{{q\pi }}-3 T_0^2 \bar{\phi } C_{{T\pi }} \epsilon_{\phi } \kappa _{{q\pi }}+3 \chi  \bar{\phi } \epsilon_{\phi }\nn\\
&-&3 T_0^2 C_{{n\pi }} \epsilon_T n_0 \kappa _{{q\pi }}+3 T_0^2 C_{{T\pi }} \epsilon_n n_0 \kappa _{{q\pi }}-3 \chi  \epsilon_n n_0-3 \epsilon_{\phi } P_0 \kappa _{{q\pi }}-3 \epsilon_0 \epsilon_{\phi } \kappa _{{q\pi }}-3 T_0 \epsilon_{\phi } P_T \kappa _{{q\pi }}\nn\\
&+&3 T_0 \epsilon_T P_{\phi } \kappa _{{q\pi }}+\zeta  \epsilon_T+4 \eta  \epsilon_T+3 T_0 \chi  P_T\Big]+\mathcal{O}(k^3)\,,
\label{A8}
\eeqa
\beqa
\omega_{4} &=&\frac{i \gamma  (C_{\phi \pi } \epsilon_T-C_{{T\pi }} e_{\phi })}{3 \epsilon_T}+k \sqrt{\frac{\bar{\phi } \epsilon_{\phi } P_T-\bar{\phi } \epsilon_T P_{\phi }+\epsilon_T n_0 P_n-\epsilon_n n_0 P_T+\epsilon_0 P_T+P_0 P_T}{\epsilon_T (\epsilon_0+P_0)}}\nn\\
&+&\frac{ik^{2}}{{9 \epsilon_T (\epsilon_0+P_0)}}\Big[(3 T_0^2 \bar{\phi } C_{\phi \pi } \epsilon_T \kappa _{{q\pi }}-3 T_0^2 \bar{\phi } C_{{T\pi }} \epsilon_{\phi } \kappa _{{q\pi }}+3 \chi  \bar{\phi } \epsilon_{\phi }-3 T_0^2 C_{{n\pi }} \epsilon_T n_0 \kappa _{{q\pi }}\nn\\
&+&3 T_0^2 C_{{T\pi }} \epsilon_n n_0 \kappa _{{q\pi }}-3 \chi  \epsilon_n n_0-3 \epsilon_{\phi } P_0 \kappa _{{q\pi }}-3 \epsilon_0 \epsilon_{\phi } \kappa _{{q\pi }}-3 T_0 \epsilon_{\phi } P_T \kappa _{{q\pi }}+3 T_0 \epsilon_T P_{\phi } \kappa _{{q\pi }}\nn\\
&+&\zeta  \epsilon_T+4 \eta  \epsilon_T+3 T_0 \chi  P_T\Big]+\mathcal{O}(k^3)\,.
\label{A9}
\eeqa
The width of the Brillouin peaks can be identified as
\beqa
\Gamma _B&=&-{9 \epsilon_T (\epsilon_0+P_0)}\Big[3 T_0^2 \bar{\phi } C_{\phi \pi } \epsilon_T \kappa _{{q\pi }}-3 T_0^2 \bar{\phi } C_{{T\pi }} \epsilon_{\phi } \kappa _{{q\pi }}+3 \chi  \bar{\phi } \epsilon_{\phi }-3 T_0^2 C_{{n\pi }} \epsilon_T n_0 \kappa _{{q\pi }}\nn\\
&+&3 T_0^2 C_{{T\pi }} \epsilon_n n_0 \kappa _{{q\pi }}-3 \chi  \epsilon_n n_0-3 \epsilon_{\phi } P_0 \kappa _{{q\pi }}-3 \epsilon_0 \epsilon_{\phi } \kappa _{{q\pi }}-3 T_0 \epsilon_{\phi } P_T \kappa _{{q\pi }}+3 T_0 \epsilon_T P_{\phi } \kappa _{{q\pi }}\nn\\
&+&\zeta  \epsilon_T+4 \eta  \epsilon_T+3 T_0 \chi  P_T\Big]\,.
\label{A10}
\eeqa
The speed of sound obtained from the dispersion relation is expressed as:
\beqa
c^{2}_s=\frac{\epsilon_T n_0 P_n-\epsilon_n n_0 P_T+\epsilon_0 P_T+P_0 P_T+\bar{\phi } \epsilon_{\phi } P_T-\bar{\phi } \epsilon_T P_{\phi }}{\epsilon_T (\epsilon_0+P_0)}\,.
\label{A11}
\eeqa
where, in Eqs.\eqref{A8}-\eqref{A11}, we have used the notation as
\beqa
X_Y=\Big(\frac{\partial X}{\partial Y}\Big)\,.
\eeqa
\bibliography{Slowmodes}

\begin{thebibliography}{44}
\expandafter\ifx\csname natexlab\endcsname\relax\def\natexlab#1{#1}\fi
\expandafter\ifx\csname bibnamefont\endcsname\relax
  \def\bibnamefont#1{#1}\fi
\expandafter\ifx\csname bibfnamefont\endcsname\relax
  \def\bibfnamefont#1{#1}\fi
\expandafter\ifx\csname citenamefont\endcsname\relax
  \def\citenamefont#1{#1}\fi
\expandafter\ifx\csname url\endcsname\relax
  \def\url#1{\texttt{#1}}\fi
\expandafter\ifx\csname urlprefix\endcsname\relax\def\urlprefix{URL }\fi
\providecommand{\bibinfo}[2]{#2}
\providecommand{\eprint}[2][]{\url{#2}}

\bibitem[{\citenamefont{Halasz et~al.}(1998)\citenamefont{Halasz, Jackson,
  Shrock, Stephanov, and Verbaarschot}}]{Halasz:1998qr}
\bibinfo{author}{\bibfnamefont{A.~M.} \bibnamefont{Halasz}},
  \bibinfo{author}{\bibfnamefont{A.~D.} \bibnamefont{Jackson}},
  \bibinfo{author}{\bibfnamefont{R.~E.} \bibnamefont{Shrock}},
  \bibinfo{author}{\bibfnamefont{M.~A.} \bibnamefont{Stephanov}},
  \bibnamefont{and} \bibinfo{author}{\bibfnamefont{J.~J.~M.}
  \bibnamefont{Verbaarschot}}, \bibinfo{journal}{Phys. Rev. D}
  \textbf{\bibinfo{volume}{58}}, \bibinfo{pages}{096007}
  (\bibinfo{year}{1998}), \eprint{hep-ph/9804290}.

\bibitem[{\citenamefont{Barducci et~al.}(1994)\citenamefont{Barducci,
  Casalbuoni, Pettini, and Gatto}}]{PhysRevD.49.426}
\bibinfo{author}{\bibfnamefont{A.}~\bibnamefont{Barducci}},
  \bibinfo{author}{\bibfnamefont{R.}~\bibnamefont{Casalbuoni}},
  \bibinfo{author}{\bibfnamefont{G.}~\bibnamefont{Pettini}}, \bibnamefont{and}
  \bibinfo{author}{\bibfnamefont{R.}~\bibnamefont{Gatto}},
  \bibinfo{journal}{Phys. Rev. D} \textbf{\bibinfo{volume}{49}},
  \bibinfo{pages}{426} (\bibinfo{year}{1994}),
  \urlprefix\url{https://link.aps.org/doi/10.1103/PhysRevD.49.426}.

\bibitem[{\citenamefont{Barducci
  et~al.}(1990{\natexlab{a}})\citenamefont{Barducci, Casalbuoni, De~Curtis,
  Gatto, and Pettini}}]{PhysRevD.42.1757}
\bibinfo{author}{\bibfnamefont{A.}~\bibnamefont{Barducci}},
  \bibinfo{author}{\bibfnamefont{R.}~\bibnamefont{Casalbuoni}},
  \bibinfo{author}{\bibfnamefont{S.}~\bibnamefont{De~Curtis}},
  \bibinfo{author}{\bibfnamefont{R.}~\bibnamefont{Gatto}}, \bibnamefont{and}
  \bibinfo{author}{\bibfnamefont{G.}~\bibnamefont{Pettini}},
  \bibinfo{journal}{Phys. Rev. D} \textbf{\bibinfo{volume}{42}},
  \bibinfo{pages}{1757} (\bibinfo{year}{1990}{\natexlab{a}}),
  \urlprefix\url{https://link.aps.org/doi/10.1103/PhysRevD.42.1757}.

\bibitem[{\citenamefont{Berges and Rajagopal}(1999)}]{BERGES1999215}
\bibinfo{author}{\bibfnamefont{J.}~\bibnamefont{Berges}} \bibnamefont{and}
  \bibinfo{author}{\bibfnamefont{K.}~\bibnamefont{Rajagopal}},
  \bibinfo{journal}{Nuclear Physics B} \textbf{\bibinfo{volume}{538}},
  \bibinfo{pages}{215} (\bibinfo{year}{1999}), ISSN \bibinfo{issn}{0550-3213},
  \urlprefix\url{https://www.sciencedirect.com/science/article/pii/S0550321398006208}.

\bibitem[{\citenamefont{Kiriyama et~al.}(2000)\citenamefont{Kiriyama, Maruyama,
  and Takagi}}]{PhysRevD.62.105008}
\bibinfo{author}{\bibfnamefont{O.}~\bibnamefont{Kiriyama}},
  \bibinfo{author}{\bibfnamefont{M.}~\bibnamefont{Maruyama}}, \bibnamefont{and}
  \bibinfo{author}{\bibfnamefont{F.}~\bibnamefont{Takagi}},
  \bibinfo{journal}{Phys. Rev. D} \textbf{\bibinfo{volume}{62}},
  \bibinfo{pages}{105008} (\bibinfo{year}{2000}),
  \urlprefix\url{https://link.aps.org/doi/10.1103/PhysRevD.62.105008}.

\bibitem[{\citenamefont{Fodor and Katz}(2002)}]{FODOR200287}
\bibinfo{author}{\bibfnamefont{Z.}~\bibnamefont{Fodor}} \bibnamefont{and}
  \bibinfo{author}{\bibfnamefont{S.}~\bibnamefont{Katz}},
  \bibinfo{journal}{Physics Letters B} \textbf{\bibinfo{volume}{534}},
  \bibinfo{pages}{87} (\bibinfo{year}{2002}), ISSN \bibinfo{issn}{0370-2693},
  \urlprefix\url{https://www.sciencedirect.com/science/article/pii/S0370269302015836}.

\bibitem[{\citenamefont{Fodor and Katz}(2004)}]{Fodor:2004nz}
\bibinfo{author}{\bibfnamefont{Z.}~\bibnamefont{Fodor}} \bibnamefont{and}
  \bibinfo{author}{\bibfnamefont{S.~D.} \bibnamefont{Katz}},
  \bibinfo{journal}{JHEP} \textbf{\bibinfo{volume}{04}}, \bibinfo{pages}{050}
  (\bibinfo{year}{2004}), \eprint{hep-lat/0402006}.

\bibitem[{\citenamefont{Aggarwal et~al.}(2010)}]{STAR:2010vob}
\bibinfo{author}{\bibfnamefont{M.~M.} \bibnamefont{Aggarwal}}
  \bibnamefont{et~al.} (\bibinfo{collaboration}{STAR}) (\bibinfo{year}{2010}),
  \eprint{1007.2613}.

\bibitem[{\citenamefont{Gavai}(2015)}]{Gavai:2014ela}
\bibinfo{author}{\bibfnamefont{R.~V.} \bibnamefont{Gavai}},
  \bibinfo{journal}{Pramana} \textbf{\bibinfo{volume}{84}},
  \bibinfo{pages}{757} (\bibinfo{year}{2015}), \eprint{1404.6615}.

\bibitem[{\citenamefont{Masayuki and Koichi}(1989)}]{MASAYUKI1989668}
\bibinfo{author}{\bibfnamefont{A.}~\bibnamefont{Masayuki}} \bibnamefont{and}
  \bibinfo{author}{\bibfnamefont{Y.}~\bibnamefont{Koichi}},
  \bibinfo{journal}{Nuclear Physics A} \textbf{\bibinfo{volume}{504}},
  \bibinfo{pages}{668} (\bibinfo{year}{1989}), ISSN \bibinfo{issn}{0375-9474},
  \urlprefix\url{https://www.sciencedirect.com/science/article/pii/037594748990002X}.

\bibitem[{\citenamefont{Barducci
  et~al.}(1990{\natexlab{b}})\citenamefont{Barducci, Casalbuoni, De~Curtis,
  Gatto, and Pettini}}]{PhysRevD.41.1610}
\bibinfo{author}{\bibfnamefont{A.}~\bibnamefont{Barducci}},
  \bibinfo{author}{\bibfnamefont{R.}~\bibnamefont{Casalbuoni}},
  \bibinfo{author}{\bibfnamefont{S.}~\bibnamefont{De~Curtis}},
  \bibinfo{author}{\bibfnamefont{R.}~\bibnamefont{Gatto}}, \bibnamefont{and}
  \bibinfo{author}{\bibfnamefont{G.}~\bibnamefont{Pettini}},
  \bibinfo{journal}{Phys. Rev. D} \textbf{\bibinfo{volume}{41}},
  \bibinfo{pages}{1610} (\bibinfo{year}{1990}{\natexlab{b}}),
  \urlprefix\url{https://link.aps.org/doi/10.1103/PhysRevD.41.1610}.

\bibitem[{\citenamefont{Du et~al.}(2021)\citenamefont{Du, An, and
  Heinz}}]{Lupeidu2021}
\bibinfo{author}{\bibfnamefont{L.}~\bibnamefont{Du}},
  \bibinfo{author}{\bibfnamefont{X.}~\bibnamefont{An}}, \bibnamefont{and}
  \bibinfo{author}{\bibfnamefont{U.}~\bibnamefont{Heinz}},
  \bibinfo{journal}{Phys. Rev. C} \textbf{\bibinfo{volume}{104}},
  \bibinfo{pages}{064904} (\bibinfo{year}{2021}),
  \urlprefix\url{https://link.aps.org/doi/10.1103/PhysRevC.104.064904}.

\bibitem[{\citenamefont{Bleicher and Herold}(2012)}]{Bleicher:2012qve}
\bibinfo{author}{\bibfnamefont{M.}~\bibnamefont{Bleicher}} \bibnamefont{and}
  \bibinfo{author}{\bibfnamefont{C.}~\bibnamefont{Herold}},
  \bibinfo{journal}{PoS} \textbf{\bibinfo{volume}{ConfinementX}},
  \bibinfo{pages}{217} (\bibinfo{year}{2012}), \eprint{1303.3686}.

\bibitem[{\citenamefont{Aguiar et~al.}(2007)\citenamefont{Aguiar, Kodama,
  Koide, and Hama}}]{Aguiar2007}
\bibinfo{author}{\bibfnamefont{C.}~\bibnamefont{Aguiar}},
  \bibinfo{author}{\bibfnamefont{T.}~\bibnamefont{Kodama}},
  \bibinfo{author}{\bibfnamefont{T.}~\bibnamefont{Koide}}, \bibnamefont{and}
  \bibinfo{author}{\bibfnamefont{Y.}~\bibnamefont{Hama}},
  \bibinfo{journal}{Brazilian Journal of Physics - BRAZ J PHYS}
  \textbf{\bibinfo{volume}{37}}, \bibinfo{pages}{95} (\bibinfo{year}{2007}).

\bibitem[{\citenamefont{Nonaka and Asakawa}(2005)}]{Nonaka}
\bibinfo{author}{\bibfnamefont{C.}~\bibnamefont{Nonaka}} \bibnamefont{and}
  \bibinfo{author}{\bibfnamefont{M.}~\bibnamefont{Asakawa}},
  \bibinfo{journal}{Phys. Rev. C} \textbf{\bibinfo{volume}{71}},
  \bibinfo{pages}{044904} (\bibinfo{year}{2005}),
  \urlprefix\url{https://link.aps.org/doi/10.1103/PhysRevC.71.044904}.

\bibitem[{\citenamefont{Stephanov and Yin}(2018)}]{Stephanov:2017ghc}
\bibinfo{author}{\bibfnamefont{M.}~\bibnamefont{Stephanov}} \bibnamefont{and}
  \bibinfo{author}{\bibfnamefont{Y.}~\bibnamefont{Yin}},
  \bibinfo{journal}{Phys. Rev. D} \textbf{\bibinfo{volume}{98}},
  \bibinfo{pages}{036006} (\bibinfo{year}{2018}),
  \urlprefix\url{https://link.aps.org/doi/10.1103/PhysRevD.98.036006}.

\bibitem[{\citenamefont{Asakawa et~al.}(2000)\citenamefont{Asakawa, Heinz, and
  M\"uller}}]{AsakawaPhysRevLett.85.2072}
\bibinfo{author}{\bibfnamefont{M.}~\bibnamefont{Asakawa}},
  \bibinfo{author}{\bibfnamefont{U.}~\bibnamefont{Heinz}}, \bibnamefont{and}
  \bibinfo{author}{\bibfnamefont{B.}~\bibnamefont{M\"uller}},
  \bibinfo{journal}{Phys. Rev. Lett.} \textbf{\bibinfo{volume}{85}},
  \bibinfo{pages}{2072} (\bibinfo{year}{2000}),
  \urlprefix\url{https://link.aps.org/doi/10.1103/PhysRevLett.85.2072}.

\bibitem[{\citenamefont{Jeon and Koch}(2003)}]{Jeon2003eventbyevent}
\bibinfo{author}{\bibfnamefont{S.}~\bibnamefont{Jeon}} \bibnamefont{and}
  \bibinfo{author}{\bibfnamefont{V.}~\bibnamefont{Koch}}
  (\bibinfo{year}{2003}), \eprint{hep-ph/0304012}.

\bibitem[{\citenamefont{Koch}(2008)}]{Koch2008hadronic}
\bibinfo{author}{\bibfnamefont{V.}~\bibnamefont{Koch}} (\bibinfo{year}{2008}),
  \eprint{0810.2520}.

\bibitem[{\citenamefont{Berdnikov and Rajagopal}(2000)}]{Berdnikov:1999ph}
\bibinfo{author}{\bibfnamefont{B.}~\bibnamefont{Berdnikov}} \bibnamefont{and}
  \bibinfo{author}{\bibfnamefont{K.}~\bibnamefont{Rajagopal}},
  \bibinfo{journal}{Phys. Rev. D} \textbf{\bibinfo{volume}{61}},
  \bibinfo{pages}{105017} (\bibinfo{year}{2000}),
  \urlprefix\url{https://link.aps.org/doi/10.1103/PhysRevD.61.105017}.

\bibitem[{\citenamefont{Stanley}(1987)}]{Stanley}
\bibinfo{author}{\bibfnamefont{H.}~\bibnamefont{Stanley}},
  \emph{\bibinfo{title}{Introduction to Phase Transitions and Critical
  Phenomena}}, International series of monographs on physics
  (\bibinfo{publisher}{Oxford University Press}, \bibinfo{year}{1987}), ISBN
  \bibinfo{isbn}{9780195053166},
  \urlprefix\url{https://books.google.co.in/books?id=C3BzcUxoaNkC}.

\bibitem[{\citenamefont{Ma}(1976)}]{SKMa}
\bibinfo{author}{\bibfnamefont{S.~K.} \bibnamefont{Ma}},
  \emph{\bibinfo{title}{Modern theory of critical phenomena}}
  (\bibinfo{publisher}{Oxford University Press}, \bibinfo{year}{1976}),
  \urlprefix\url{https://www.osti.gov/biblio/7362153}.

\bibitem[{\citenamefont{Rajagopal et~al.}(2020)\citenamefont{Rajagopal,
  Ridgway, Weller, and Yin}}]{Rajagopal2020}
\bibinfo{author}{\bibfnamefont{K.}~\bibnamefont{Rajagopal}},
  \bibinfo{author}{\bibfnamefont{G.~W.} \bibnamefont{Ridgway}},
  \bibinfo{author}{\bibfnamefont{R.}~\bibnamefont{Weller}}, \bibnamefont{and}
  \bibinfo{author}{\bibfnamefont{Y.}~\bibnamefont{Yin}},
  \bibinfo{journal}{Phys. Rev. D} \textbf{\bibinfo{volume}{102}},
  \bibinfo{pages}{094025} (\bibinfo{year}{2020}),
  \urlprefix\url{https://link.aps.org/doi/10.1103/PhysRevD.102.094025}.

\bibitem[{\citenamefont{Israel}(1976)}]{Israel:1976tn}
\bibinfo{author}{\bibfnamefont{W.}~\bibnamefont{Israel}},
  \bibinfo{journal}{Annals Phys.} \textbf{\bibinfo{volume}{100}},
  \bibinfo{pages}{310} (\bibinfo{year}{1976}).

\bibitem[{\citenamefont{Ma}(2018)}]{SKMA2018}
\bibinfo{author}{\bibfnamefont{S.~K.} \bibnamefont{Ma}},
  \emph{\bibinfo{title}{Modern Theory Of Critical Phenomena}}
  (\bibinfo{publisher}{Taylor \& Francis}, \bibinfo{year}{2018}), ISBN
  \bibinfo{isbn}{9780429967436},
  \urlprefix\url{https://books.google.co.in/books?id=t8TADwAAQBAJ}.

\bibitem[{\citenamefont{E.~Reichl}()}]{Linda}
\bibinfo{author}{\bibfnamefont{L.}~\bibnamefont{E.~Reichl}},
  \emph{\bibinfo{title}{A Modern Course in Statistical Physics}} (????).

\bibitem[{\citenamefont{Onsager}(1931)}]{OnsagarPhysRev.37.405}
\bibinfo{author}{\bibfnamefont{L.}~\bibnamefont{Onsager}},
  \bibinfo{journal}{Phys. Rev.} \textbf{\bibinfo{volume}{37}},
  \bibinfo{pages}{405} (\bibinfo{year}{1931}),
  \urlprefix\url{https://link.aps.org/doi/10.1103/PhysRev.37.405}.

\bibitem[{\citenamefont{Rayleigh}(1881)}]{Rayleigh1881}
\bibinfo{author}{\bibfnamefont{L.}~\bibnamefont{Rayleigh}},
  \bibinfo{journal}{The London, Edinburgh, and Dublin Philosophical Magazine
  and Journal of Science} \textbf{\bibinfo{volume}{12}}, \bibinfo{pages}{81}
  (\bibinfo{year}{1881}), \eprint{https://doi.org/10.1080/14786448108627074},
  \urlprefix\url{https://doi.org/10.1080/14786448108627074}.

\bibitem[{\citenamefont{Fleury and Boon}(1969)}]{FlerryandBoon1969}
\bibinfo{author}{\bibfnamefont{P.~A.} \bibnamefont{Fleury}} \bibnamefont{and}
  \bibinfo{author}{\bibfnamefont{J.~P.} \bibnamefont{Boon}},
  \bibinfo{journal}{Phys. Rev.} \textbf{\bibinfo{volume}{186}},
  \bibinfo{pages}{244} (\bibinfo{year}{1969}),
  \urlprefix\url{https://link.aps.org/doi/10.1103/PhysRev.186.244}.

\bibitem[{\citenamefont{Minami and Kunihiro}(2009)}]{10.1143/PTP.122.881}
\bibinfo{author}{\bibfnamefont{Y.}~\bibnamefont{Minami}} \bibnamefont{and}
  \bibinfo{author}{\bibfnamefont{T.}~\bibnamefont{Kunihiro}},
  \bibinfo{journal}{Progress of Theoretical Physics}
  \textbf{\bibinfo{volume}{122}}, \bibinfo{pages}{881} (\bibinfo{year}{2009}),
  ISSN \bibinfo{issn}{0033-068X},
  \eprint{https://academic.oup.com/ptp/article-pdf/122/4/881/9681298/122-4-881.pdf},
  \urlprefix\url{https://doi.org/10.1143/PTP.122.881}.

\bibitem[{\citenamefont{Minami and Kunihiro}(2010)}]{Minami}
\bibinfo{author}{\bibfnamefont{Y.}~\bibnamefont{Minami}} \bibnamefont{and}
  \bibinfo{author}{\bibfnamefont{T.}~\bibnamefont{Kunihiro}},
  \bibinfo{journal}{Prog. Theor. Phys.} \textbf{\bibinfo{volume}{122}},
  \bibinfo{pages}{881} (\bibinfo{year}{2010}), \eprint{0904.2270}.

\bibitem[{\citenamefont{Hasanujjaman et~al.}(2021)\citenamefont{Hasanujjaman,
  Sarwar, Rahaman, Bhattacharyya, and Alam}}]{Hasan2}
\bibinfo{author}{\bibfnamefont{M.}~\bibnamefont{Hasanujjaman}},
  \bibinfo{author}{\bibfnamefont{G.}~\bibnamefont{Sarwar}},
  \bibinfo{author}{\bibfnamefont{M.}~\bibnamefont{Rahaman}},
  \bibinfo{author}{\bibfnamefont{A.}~\bibnamefont{Bhattacharyya}},
  \bibnamefont{and} \bibinfo{author}{\bibfnamefont{J.-e.} \bibnamefont{Alam}},
  \bibinfo{journal}{Eur. Phys. J. A} \textbf{\bibinfo{volume}{57}},
  \bibinfo{pages}{283} (\bibinfo{year}{2021}), \eprint{2008.03931}.

\bibitem[{\citenamefont{Nagai et~al.}(2016)\citenamefont{Nagai, Kurita, Murase,
  and Hirano}}]{Nagai2016}
\bibinfo{author}{\bibfnamefont{K.}~\bibnamefont{Nagai}},
  \bibinfo{author}{\bibfnamefont{R.}~\bibnamefont{Kurita}},
  \bibinfo{author}{\bibfnamefont{K.}~\bibnamefont{Murase}}, \bibnamefont{and}
  \bibinfo{author}{\bibfnamefont{T.}~\bibnamefont{Hirano}},
  \bibinfo{journal}{Nucl. Phys. A} \textbf{\bibinfo{volume}{956}},
  \bibinfo{pages}{781} (\bibinfo{year}{2016}), \eprint{1602.00794}.

\bibitem[{\citenamefont{Chatrchyan et~al.}(2013)}]{CMS:2013jlh}
\bibinfo{author}{\bibfnamefont{S.}~\bibnamefont{Chatrchyan}}
  \bibnamefont{et~al.} (\bibinfo{collaboration}{CMS}), \bibinfo{journal}{Phys.
  Lett. B} \textbf{\bibinfo{volume}{724}}, \bibinfo{pages}{213}
  (\bibinfo{year}{2013}), \eprint{1305.0609}.

\bibitem[{\citenamefont{Eckart}(1940)}]{Eckart:1940te}
\bibinfo{author}{\bibfnamefont{C.}~\bibnamefont{Eckart}},
  \bibinfo{journal}{Phys. Rev.} \textbf{\bibinfo{volume}{58}},
  \bibinfo{pages}{919} (\bibinfo{year}{1940}).

\bibitem[{\citenamefont{Muronga}(2004)}]{Muronga:2003ta}
\bibinfo{author}{\bibfnamefont{A.}~\bibnamefont{Muronga}},
  \bibinfo{journal}{Phys. Rev. C} \textbf{\bibinfo{volume}{69}},
  \bibinfo{pages}{034903} (\bibinfo{year}{2004}), \eprint{nucl-th/0309055}.

\bibitem[{\citenamefont{Muronga}(2002)}]{Muronga:2001zk}
\bibinfo{author}{\bibfnamefont{A.}~\bibnamefont{Muronga}},
  \bibinfo{journal}{Phys. Rev. Lett.} \textbf{\bibinfo{volume}{88}},
  \bibinfo{pages}{062302} (\bibinfo{year}{2002}), \bibinfo{note}{[Erratum:
  Phys.Rev.Lett. 89, 159901 (2002)]}, \eprint{nucl-th/0104064}.

\bibitem[{\citenamefont{Parotto et~al.}(2020)\citenamefont{Parotto, Bluhm,
  Mroczek, Nahrgang, Noronha-Hostler, Rajagopal, Ratti, Sch\"afer, and
  Stephanov}}]{Parotto}
\bibinfo{author}{\bibfnamefont{P.}~\bibnamefont{Parotto}},
  \bibinfo{author}{\bibfnamefont{M.}~\bibnamefont{Bluhm}},
  \bibinfo{author}{\bibfnamefont{D.}~\bibnamefont{Mroczek}},
  \bibinfo{author}{\bibfnamefont{M.}~\bibnamefont{Nahrgang}},
  \bibinfo{author}{\bibfnamefont{J.}~\bibnamefont{Noronha-Hostler}},
  \bibinfo{author}{\bibfnamefont{K.}~\bibnamefont{Rajagopal}},
  \bibinfo{author}{\bibfnamefont{C.}~\bibnamefont{Ratti}},
  \bibinfo{author}{\bibfnamefont{T.}~\bibnamefont{Sch\"afer}},
  \bibnamefont{and}
  \bibinfo{author}{\bibfnamefont{M.}~\bibnamefont{Stephanov}},
  \bibinfo{journal}{Phys. Rev. C} \textbf{\bibinfo{volume}{101}},
  \bibinfo{pages}{034901} (\bibinfo{year}{2020}),
  \urlprefix\url{https://link.aps.org/doi/10.1103/PhysRevC.101.034901}.

\bibitem[{\citenamefont{Hasanujjaman et~al.}(2020)\citenamefont{Hasanujjaman,
  Rahaman, Bhattacharyya, and Alam}}]{Hasan1}
\bibinfo{author}{\bibfnamefont{M.}~\bibnamefont{Hasanujjaman}},
  \bibinfo{author}{\bibfnamefont{M.}~\bibnamefont{Rahaman}},
  \bibinfo{author}{\bibfnamefont{A.}~\bibnamefont{Bhattacharyya}},
  \bibnamefont{and} \bibinfo{author}{\bibfnamefont{J.-e.} \bibnamefont{Alam}},
  \bibinfo{journal}{Phys. Rev. C} \textbf{\bibinfo{volume}{102}},
  \bibinfo{pages}{034910} (\bibinfo{year}{2020}), \eprint{2003.07575}.

\bibitem[{\citenamefont{Kapusta and Torres-Rincon}(2012)}]{KapustaChi}
\bibinfo{author}{\bibfnamefont{J.~I.} \bibnamefont{Kapusta}} \bibnamefont{and}
  \bibinfo{author}{\bibfnamefont{J.~M.} \bibnamefont{Torres-Rincon}},
  \bibinfo{journal}{Phys. Rev. C} \textbf{\bibinfo{volume}{86}},
  \bibinfo{pages}{054911} (\bibinfo{year}{2012}),
  \urlprefix\url{https://link.aps.org/doi/10.1103/PhysRevC.86.054911}.

\bibitem[{\citenamefont{Guida and Zinn-Justin}(1997)}]{Guida}
\bibinfo{author}{\bibfnamefont{R.}~\bibnamefont{Guida}} \bibnamefont{and}
  \bibinfo{author}{\bibfnamefont{J.}~\bibnamefont{Zinn-Justin}},
  \bibinfo{journal}{Nuclear Physics B} \textbf{\bibinfo{volume}{489}},
  \bibinfo{pages}{626} (\bibinfo{year}{1997}), ISSN \bibinfo{issn}{0550-3213},
  \urlprefix\url{https://www.sciencedirect.com/science/article/pii/S0550321396007043}.

\bibitem[{\citenamefont{Rajagopal and Wilczek}(1993)}]{Rajagopal:1992qz}
\bibinfo{author}{\bibfnamefont{K.}~\bibnamefont{Rajagopal}} \bibnamefont{and}
  \bibinfo{author}{\bibfnamefont{F.}~\bibnamefont{Wilczek}},
  \bibinfo{journal}{Nucl. Phys. B} \textbf{\bibinfo{volume}{399}},
  \bibinfo{pages}{395} (\bibinfo{year}{1993}), \eprint{hep-ph/9210253}.

\bibitem[{\citenamefont{Rayleigh}(1916)}]{Rayeligh_Benard}
\bibinfo{author}{\bibfnamefont{L.}~\bibnamefont{Rayleigh}},
  \bibinfo{journal}{The London, Edinburgh, and Dublin Philosophical Magazine
  and Journal of Science} \textbf{\bibinfo{volume}{32}}, \bibinfo{pages}{529}
  (\bibinfo{year}{1916}), \eprint{https://doi.org/10.1080/14786441608635602},
  \urlprefix\url{https://doi.org/10.1080/14786441608635602}.

\bibitem[{\citenamefont{Zarate and Sengers}(2006)}]{Zarate}
\bibinfo{author}{\bibfnamefont{J.}~\bibnamefont{Zarate}} \bibnamefont{and}
  \bibinfo{author}{\bibfnamefont{J.}~\bibnamefont{Sengers}},
  \bibinfo{journal}{Hydrodynamic Fluctuations in Fluids and Fluid Mixtures}
  (\bibinfo{year}{2006}).

\end{thebibliography}
 
\end{document}